\documentclass{article}
\usepackage{authblk}
\usepackage{float}
\usepackage{graphicx}
\usepackage{hyperref}
\usepackage{listings}
\usepackage{pgfplots}
\usepackage{tikz}
\usepackage{xcolor}
\usetikzlibrary{shapes.geometric, arrows}

\definecolor{codegreen}{rgb}{0,0.6,0}
\definecolor{codegray}{rgb}{0.5,0.5,0.5}
\definecolor{codepurple}{rgb}{0.58,0,0.82}
\definecolor{backcolour}{rgb}{0.95,0.95,0.92}

\lstdefinestyle{mystyle} {
    backgroundcolor=\color{backcolour},   
    commentstyle=\color{codegreen},
    keywordstyle=\color{magenta},
    numberstyle=\tiny\color{codegray},
    stringstyle=\color{codepurple},
    basicstyle=\ttfamily\footnotesize,
    breakatwhitespace=false,         
    breaklines=true,           
    captionpos=b,                    
    keepspaces=true,                 
    numbers=left,                    
    numbersep=5pt,                  
    showspaces=false,                
    showstringspaces=false,
    showtabs=false,                  
    tabsize=2
}

\tikzstyle{startstop} = [rectangle, rounded corners, minimum width=3cm, minimum height=1cm,text centered, draw=black, fill=gray!30]
\tikzstyle{BegEnd} = [rectangle, rounded corners, minimum width=1cm, minimum height=1cm,text centered, draw=white]
\tikzstyle{realtimeds} = [rectangle, rounded corners, minimum width=5cm, minimum height=5cm, text centered, draw=black]
\tikzstyle{textbox}    = [rectangle, rounded corners, minimum width=2cm, minimum height=1cm, text centered, draw=white]
\tikzstyle{focus} = [rectangle, rounded corners, minimum width=3cm, minimum height=1cm,text centered, draw=black, fill=red!30]
\tikzstyle{arrow} = [thick,->,>=stealth]

\newcounter{example}[section]
\newenvironment{example}[1][]{\refstepcounter{example}\par\medskip
   \noindent \textbf{Example~\theexample. #1} \rmfamily}{\medskip}
\title{%
  Machine Learning Experiences \\
  \large A story of learning AI for use in enterprise \\
    software testing that can be used by anyone}
\author[1]{Michael Cohoon}
\author[2]{Debbie Furman}
\affil[1]{IBM Systems}
\affil[2]{IBM Systems}

\date{June 2025}
\pgfplotsset{compat=1.18}

\begin{document}

\maketitle

\section{Abstract}
This paper details the machine learning (ML) journey of a group of people focused on software testing. It tells the story of how this group progressed through a ML workflow (similar to the \href{https://en.wikipedia.org/wiki/Cross-industry_standard_process_for_data_mining}{CRISP-DM} process). This workflow consists of the following steps and can be used by anyone applying ML techniques to a project: gather the data; clean the data; perform feature engineering on the data; splitting the data into two sets, one for training and one for testing; choosing a machine learning model; training the model; testing the model and evaluating the model performance. By following this workflow, anyone can effectively apply ML to any project that they are doing.

\section{Why You Should Read This Paper}
We are a group working in the software testing space and we were curious as to how machine learning could apply to our roles.  We decided to write this paper to share our journey. We started by defining our problem statement and then we identified what data existed that would help us. Along the way we learned about different machine learning algorithms and we figured out how to model our data to fit into these algorithms. We continuously learned what worked and what did not work for our problem statement.

We are not experts in the field of machine learning. We are not trained data scientists. We are experts in the domain of software testing. We applied this workflow to our domain, however this workflow can be applied to any domain (ie. a domain other than software testing), and therefore any learning project. As we were educating ourselves on machine learning, we came across terms that we never heard before, whether they were statistical terms or machine learning terms. In this paper we will introduce these terms and attempt to define them in a simple to understand way, giving you an understanding of terminology that we did not know or understand at the beginning of our journey.

If you too are in the early stages of learning about machine learning techniques, then this paper is for you. We will identify pitfalls that we ran into and successes to help you in your own journey.

The following figure (figure \ref{fig:workflow1}) is a workflow of the steps needed to apply machine learning to any project.  We will start by describing the activity of gathering data.  Next we will move on to describe what it means to clean the data.  Part of cleaning the data is something called feature engineering.  We will then touch upon how and why you will need to split the data into a training and a test set.  The next step is to explore the available models and choose one to experiment with your dataset.  Once a model is chosen, it will need to be trained using your specific data.  After the model is trained, then we talk about testing the model and evaluating the model performance.  It will be useful to keep the following ML process flow (see figure \ref{fig:workflow1}) in mind to understand the rest of the paper.

\section{Disclaimer}
In this paper, we will demonstrate our journey through understanding the basics in machine learning by sharing resources, methods, ideas, and code snippets. Please understand that these examples are not endorsements. There may be alternative or better ways that we do not discuss. If you are beginning your machine learning journey, we hope that this paper is but one resource you can learn from.

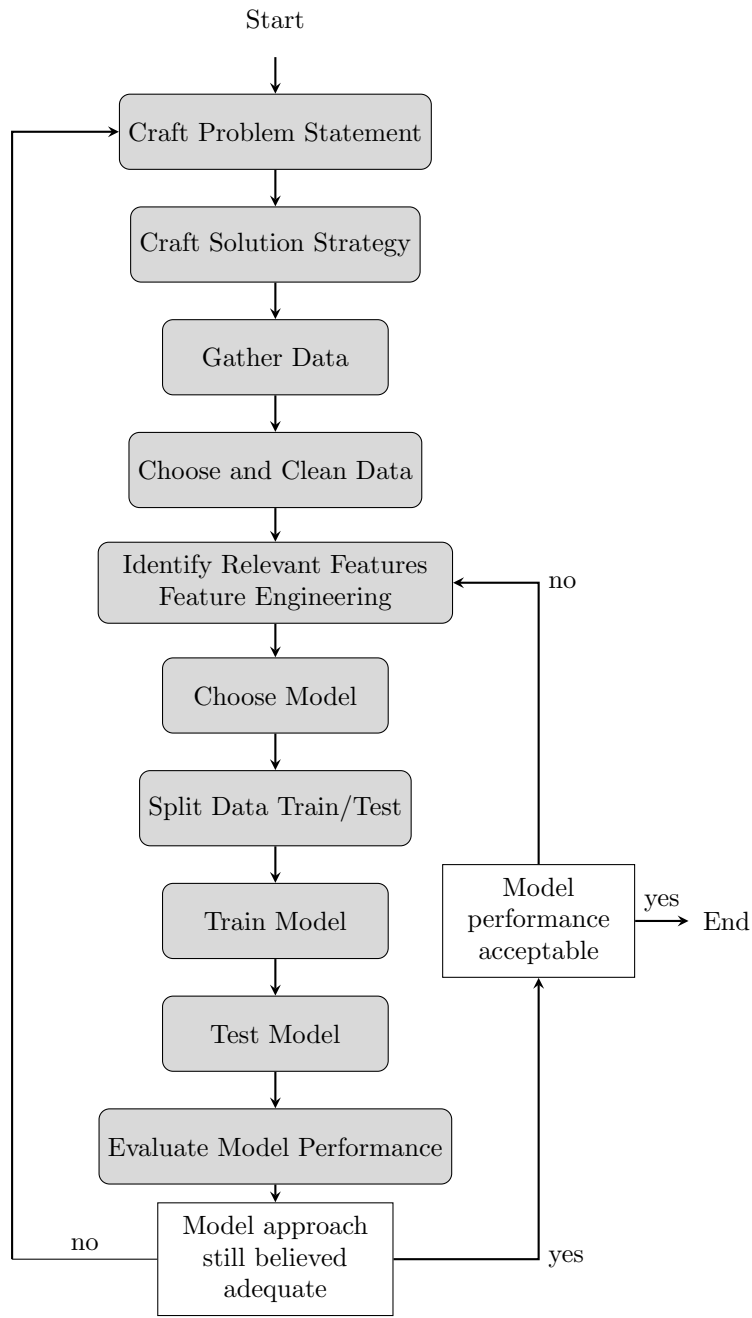
\begin{figure} [H]
  \begin{tikzpicture} [node distance=1.5cm]
    \node(Start) [BegEnd] {Start};
    \node(PS) [startstop, below of=Start] {Craft Problem Statement};
    \node(CS) [startstop, below of=PS] {Craft Solution Strategy};
    \node(GatherData) [startstop, below of=CS] {Gather Data};
    \node(ChooseData) [startstop, below of=GatherData] {Choose and Clean Data};
    \node(Feature) [startstop, below of=ChooseData]{
      \begin{tabular}{c}
      Identify Relevant Features  \\
      Feature Engineering 
      \end{tabular}
      };

    \node(ChooseModel) [startstop, below of=Feature] {Choose Model};
    \node(SplitData) [startstop, below of=ChooseModel] {Split Data Train/Test};
    \node(TrainModel) [startstop, below of=SplitData] {Train Model};

    \node(TestModel) [startstop, below of=TrainModel] {Test Model};
    \node(EP) [startstop, below of=TestModel] {Evaluate Model Performance};
    \node(DMG) [draw, below of=EP]{
      \begin{tabular}{c}
      Model approach  \\
      still believed \\
      adequate
      \end{tabular}
      };
    \node(DMF) [draw, right of=TrainModel, xshift=2cm]{
      \begin{tabular}{c}
      Model \\
      performance \\
      acceptable
      \end{tabular}
      };
    \node(End) [BegEnd, right of=DMF,xshift=1cm] {End};  
    \node(blank) [BegEnd, left of=DMG, xshift=-2cm] { };
    \draw [arrow] (Start) -- (PS);
    \draw [arrow] (PS) -- (CS);
    \draw [arrow] (CS) -- (GatherData);
    \draw [arrow] (GatherData) -- (ChooseData);
    \draw [arrow] (ChooseData) -- (Feature);
    \draw [arrow] (Feature) -- (ChooseModel);
    \draw [arrow] (ChooseModel) -- (SplitData);
    \draw [arrow] (SplitData) -- (TrainModel);
    \draw [arrow] (TrainModel) -- (TestModel);
    \draw [arrow] (TestModel) -- (EP);
    \draw [arrow] (EP) -- (DMG);
    \draw [-] (DMG) to node[anchor=south] {no} (blank.center);
    \draw [arrow] (blank.center) |- (PS);
    \draw [arrow] (DMF) |- node[anchor=west] {no} (Feature);
    \draw [arrow] (DMG) -| node[anchor=west] {yes} (DMF);
    \draw [arrow] (DMF) -- node[anchor=south] {yes} (End);
  \end{tikzpicture}
  \caption{Machine Learning Process Flow}
  \label{fig:workflow1}
\end{figure}

\newpage
\section{Problem Statement}
As software testers, we started by asking: what problems do we currently have? One of the problems that is challenging and complex is handling defects that are found both during the development cycle and in the field. For the field defects, we wondered if we could use machine learning to help predict when a fix for a field defect would have an error in it. For the defects found during the development cycle, we wondered if we could use machine learning to prune out duplicate or invalid defects.

Throughout this paper, we will be referencing both these problem statements to give examples of how we approached our education of machine learning techniques.  

We start our journey with the first lesson: namely, that machine learning is all about the data.

\section{It's all about the Data}

\begin{figure} [H]
  \begin{tikzpicture} [node distance=1.5cm]
    \node(Start) [BegEnd] {Start};
    \node(PS) [startstop, below of=Start] {Craft Problem Statement};
    \node(CS) [startstop, below of=PS] {Craft Solution Strategy};
    \node(GatherData) [focus, below of=CS] {Gather Data};
    \node(ChooseData) [startstop, below of=GatherData] {Choose and Clean Data};
    \node(Feature) [startstop, below of=ChooseData]{
      \begin{tabular}{c}
      Identify Relevant Features  \\
      Feature Engineering 
      \end{tabular}
      };
    \node(ChooseModel) [startstop, below of=Feature] {Choose Model};
    \node(SplitData) [startstop, below of=ChooseModel] {Split Data Train/Test};
    \node(TrainModel) [startstop, below of=SplitData] {Train Model};
    \node(TestModel) [startstop, below of=TrainModel] {Test Model};
    \node(EP) [startstop, below of=TestModel] {Evaluate Model Performance};
    \node(DMG) [draw, below of=EP]{
      \begin{tabular}{c}
      Model approach  \\
      still believed \\
      adequate
      \end{tabular}
      };
    \node(DMF) [draw, right of=TrainModel, xshift=2cm]{
      \begin{tabular}{c}
      Model \\
      performance \\
      acceptable
      \end{tabular}
      };
    \node(End) [BegEnd, right of=DMF,xshift=1cm] {End};  
    \node(blank) [BegEnd, left of=DMG, xshift=-2cm] { };
    \draw [arrow] (Start) -- (PS);
    \draw [arrow] (PS) -- (CS);
    \draw [arrow] (CS) -- (GatherData);
    \draw [arrow] (GatherData) -- (ChooseData);
    \draw [arrow] (ChooseData) -- (Feature);
    \draw [arrow] (Feature) -- (ChooseModel);
    \draw [arrow] (SplitData) -- (TrainModel);
    \draw [arrow] (ChooseModel) -- (SplitData);
    \draw [arrow] (TrainModel) -- (TestModel);
    \draw [arrow] (TestModel) -- (EP);
    \draw [arrow] (EP) -- (DMG);
    \draw [-] (DMG) to node[anchor=south] {no} (blank.center);
    \draw [arrow] (blank.center) |- (PS);
    \draw [arrow] (DMF) |- node[anchor=west] {no} (Feature);
    \draw [arrow] (DMG) -| node[anchor=west] {yes} (DMF);
    \draw [arrow] (DMF) -- node[anchor=south] {yes} (End);
  \end{tikzpicture}
  \caption{Gather Data}
  \label{fig:workflow2}
\end{figure}
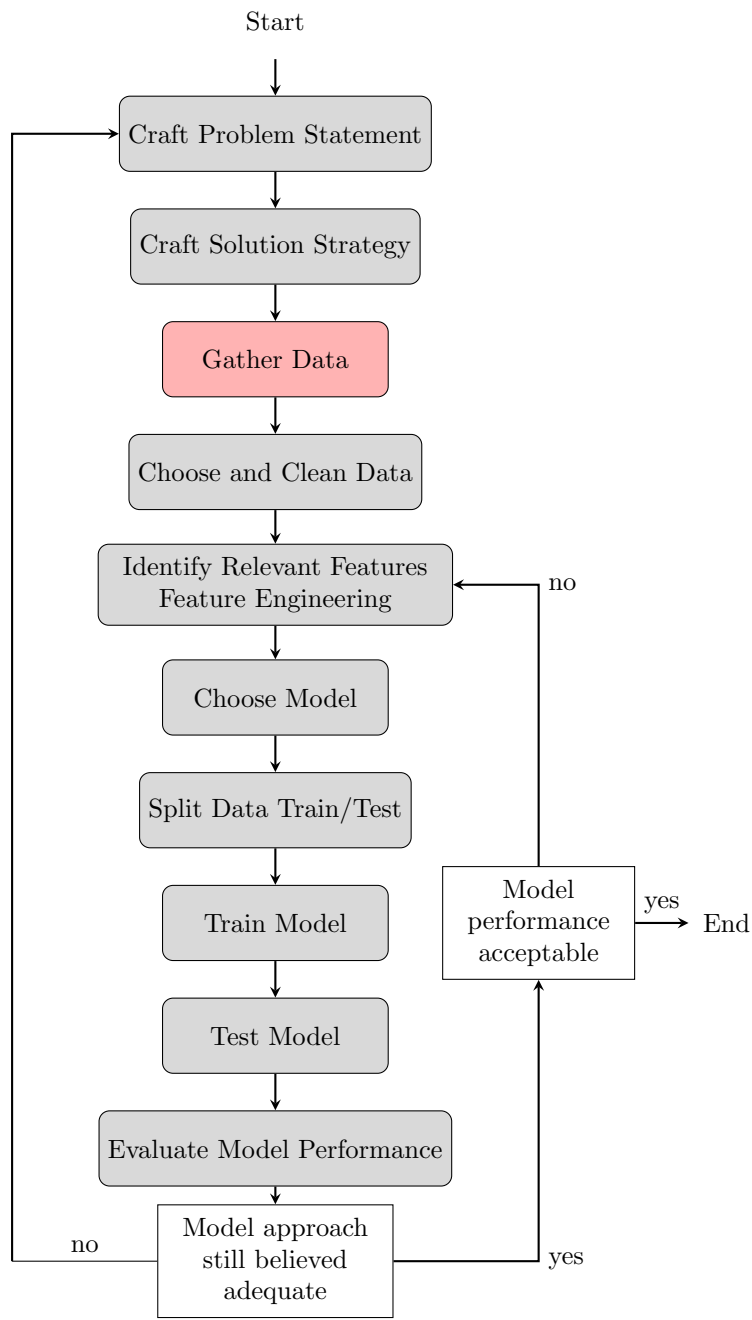

Let's look at the "Gather Data" stage in the ML process flow (see figure \ref{fig:workflow2}). 

For the application of machine learning, it is important to have both a problem to solve (or a business objective) and the data to assist in solving that problem.
In figure \ref{fig:workflow2} we show a workflow of steps we used in our machine learning journey, from identifying the problem we wanted to solve to moving through different stages of manipulating the data needed to assist our use of machine learning to solve the problem.
We first asked questions around what problems we currently had in the testing space that we would like to solve. This defined our business objective.  We then determined what data was available to help us solve these problems. For the use cases that we will go through in this paper, we have defect data both from the field after the software has been released to customers and defect data that is gathered during the software development cycle prior to being released to customers.

We say that it is important to have both a problem and data. Even if we have a problem to solve, machine learning is impossible if we don't have the data that supports the problem being looked at.  This is because machine learning requires a large amount of data in order to train the machine learning algorithm to solve a problem.  That is why we say it is all about the data.

\begin{example}
Suppose we have a business objective of predicting if a code fix needs special testing or review. It would not be valuable to answer this question with machine learning if we did not have labeled data that captures whether or not a previous code fix introduced an issue.  
\end{example}

While it is important to have the appropriate data to build a machine learning system, it is often the case that data exists without a known business objective. Have you ever worked on a software project where you
were asked to collect data without any knowledge of how the data was going to be used?  In the software testing space, we collect information regarding defects and fixes, product documentation, run-logs (execution steps and results from tests), test activities, system metrics, and more. It is possible to 
identify a business objective that requires learning from all or some of this data, but it is first important to examine the data itself.

Knowing that data exists is the first step, but often using that data is a task in itself. Discovering where
the data is housed, finding the data owner, requesting access, and exporting the data to a common platform used for machine learning are all necessary steps that are more difficult than they sound. When the data source is one that is continuously growing and updating, it drives further questions:
\begin{itemize}
    \item Should an automated process be developed to handle extracting the data?
    \item How often should the data be exported to refresh the training dataset?
\end{itemize}

\begin{figure}  
  \begin{tikzpicture} [node distance=5cm]
    \node(AA) [realtimeds]{
      \begin{tabular}{l}
      real time dataset  \\
       \\
       \\
       \\
       \\
       \\
       \\
       \\
      verification set
      \end{tabular}
    };
    \node(DD) [textbox, left of =AA, xshift=1cm ] {
      \begin{tabular}{c}
      time 1  \\
       \\
       \\
       \\
       \\
       \\
       \\
       \\
      time n \\
      \\
      \\
      \\
      time n+30
      \end{tabular}
    };
    \draw (-2.5,-1) -- (2.5,-1);
    \node(CC) [realtimeds, right of=AA, xshift=2cm]{
      \begin{tabular}{c}
       Snap shot time 1 - time n \\
       \\
       \\
       \\
       Training\\
       \\
       \\
       \\
      Testing
      \end{tabular}
    };
    \draw (4.5,-1) -- (9.5,-1);  
  \end{tikzpicture}
  \caption{Pulling snap shot of data from real time dataset}
  \label{fig:TrainTest}
\end{figure}
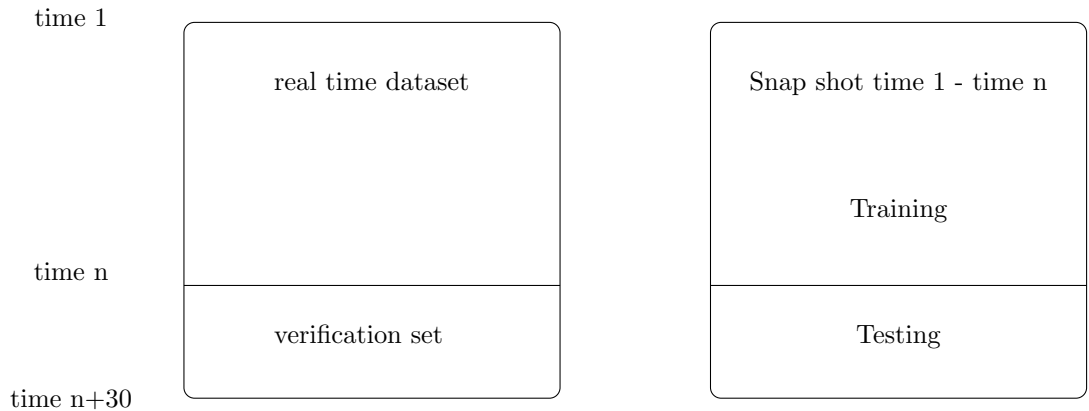

Our experience has shown that it is worth taking the time to pragmatically extract a point-in-time snapshot of the data and save it locally. Keeping a point-in-time snapshot of the data has proven beneficial for testing machine learning models and combating data drift. Splitting the point-in-time snapshot of the data into training and testing datasets helps to build a generalized machine learning model, i.e., works in new unseen cases (see figure \ref{fig:TrainTest}). The newer data, outside the point-in-time snapshot, can be used as another method of verifying the machine learning model's performance. When verifying the machine learning model's performance, we can ask ourselves, "how many desired patterns would I have detected/predicted had this machine learning model been in place?"

After we acquired access to the point-in-time snapshot of the dataset, it was important for us to review the fields in dataset. Unfortunately, our first steps were overly eager. We quickly chose existing ML packages to create a machine learning model and threw the data at it "just to see what would happen". Our results were fraught with poor and/or confusing outcomes forcing us to slow down and review the data and its features (see example \ref{def:feature} of what a feature is), something that we should have done in the first place. 

\begin{example}
    A feature represents a data point in the dataset. For example, suppose we had a dataset that represented a person. A feature of that dataset may be the person's name, address, or phone number. Each feature of the dataset would then have a value assigned to it for each record in the dataset.
    \label{def:feature}
\end{example}

Our experience shows when not being an expert on the entire dataset, it is critical to review the dataset's fields with a subject matter expert (SME). Data is typically not captured specifically for the machine learning project that you have in mind. As a result of the data not being captured for the purpose of the machine learning project, it is important to review and adjust the data as needed. 
 
\begin{example}
If our business objective is to predict valid defects based off of software defect data, we may be inclined to define features for the machine learning model for every field in the defect record. However, if some fields are optional and data is often missing for these fields, then we may need to engineer new features that identify whether or not that optional field was given a value.
Additionally, if certain fields are only used by a particular development team, then using those fields may be redundant, unnecessary, or introduce unintentional bias.
\end{example}

After working with the initial dataset, we continued discussions with the data SMEs to identify any sources of supplementary data that would help clarify fields in the initial dataset. At this point, we did not know if getting additional data would help our understanding of the original dataset, but we hoped that there would be additional benefits to learning more about the data points. For example, a data point may include different metadata of a file (such as time of creation) but finding out that a specific file resides in some part of the name-space hierarchy may be another parameter that can be utilized in the learning associated with the file. Finding any new information about the data, even if that information is external to any associated metadata, can provide valuable insights into effectively using it.

\subsection*{Lessons learned about gathering data}

Here is a cheat sheet that captures our lessons learned.

\begin{enumerate}
    \item Make sure you are allowed to use the data and identify an SME that will help you understand it. 
      \begin{itemize}
        \item Identify the owner of data and obtain the level of authority needed to perform your data cleansing.
        \item Resolve any security control issues.
      \end{itemize}
    \item Determine how often you want to refresh the data. Consider the following. 
      \begin{itemize}
          \item Do you want to refresh the model training data?
          \item Do you want to refresh the model testing data?
          \item When refreshing the data, avoid large changes in model characteristics (i.e., multiple variables changing).
      \end{itemize}
    \item Ensure the number of records in the dataset is sufficient (a common rule of thumb is to have 10x the number of records than features, but this may depend on what the data is being used for).
    \item Filter the data to make it fit to the learning task at hand.  Determine if more or less data is useful or needed.
      \begin{itemize}
          \item More data does not necessarily mean better as the additional data may include noise or biased information that is less relevant to the problem.
          \begin{itemize}
              \item In our case we were looking for defects on one platform, yet the dataset had multiple platforms. When we removed the data for the other platforms, it improved accuracy and performance of the machine learning model.
          \end{itemize}
          \item Obtaining additional data may be costly and the return on investment not sufficient in order to peruse it.
          \item May need to scope data to the business case that is being solved.
       \end{itemize}

\end{enumerate}

\newpage
\subsection{Cleaning up the Data}

\begin{figure} [H]
  \begin{tikzpicture} [node distance=1.5cm]
    \node(Start) [BegEnd] {Start};
    \node(PS) [startstop, below of=Start] {Craft Problem Statement};
    \node(CS) [startstop, below of=PS] {Craft Solution Strategy};
    \node(GatherData) [startstop, below of=CS] {Gather Data};
    \node(ChooseData) [focus, below of=GatherData] {Choose and Clean Data};
    \node(Feature) [startstop, below of=ChooseData]{
      \begin{tabular}{c}
      Identify Relevant Features  \\
      Feature Engineering 
      \end{tabular}
      };
    \node(ChooseModel) [startstop, below of=Feature] {Choose Model};
    \node(SplitData) [startstop, below of=ChooseModel] {Split Data Train/Test};
    \node(TrainModel) [startstop, below of=SplitData] {Train Model};

    \node(TestModel) [startstop, below of=TrainModel] {Test Model};
    \node(EP) [startstop, below of=TestModel] {Evaluate Model Performance};
    \node(DMG) [draw, below of=EP]{
      \begin{tabular}{c}
      Model approach  \\
      still believed \\
      adequate
      \end{tabular}
      };
    \node(DMF) [draw, right of=TrainModel, xshift=2cm]{
      \begin{tabular}{c}
      Model \\
      performance \\
      acceptable
      \end{tabular}
      };
    \node(End) [BegEnd, right of=DMF,xshift=1cm] {End};  
    \node(blank) [BegEnd, left of=DMG, xshift=-2cm] { };
    \draw [arrow] (Start) -- (PS);
    \draw [arrow] (PS) -- (CS);
    \draw [arrow] (CS) -- (GatherData);
    \draw [arrow] (GatherData) -- (ChooseData);
    \draw [arrow] (ChooseData) -- (Feature);
    \draw [arrow] (Feature) -- (ChooseModel);
    \draw [arrow] (ChooseModel) -- (SplitData);
    \draw [arrow] (SplitData) -- (TrainModel);
    \draw [arrow] (TrainModel) -- (TestModel);
    \draw [arrow] (TestModel) -- (EP);
    \draw [arrow] (EP) -- (DMG);
    \draw [-] (DMG) to node[anchor=south] {no} (blank.center);
    \draw [arrow] (blank.center) |- (PS);
    \draw [arrow] (DMF) |- node[anchor=west] {no} (Feature);
    \draw [arrow] (DMG) -| node[anchor=west] {yes} (DMF);
    \draw [arrow] (DMF) -- node[anchor=south] {yes} (End);
  \end{tikzpicture}
  \caption{Choose and Clean Data}
  \label{fig:workflow3}
\end{figure}
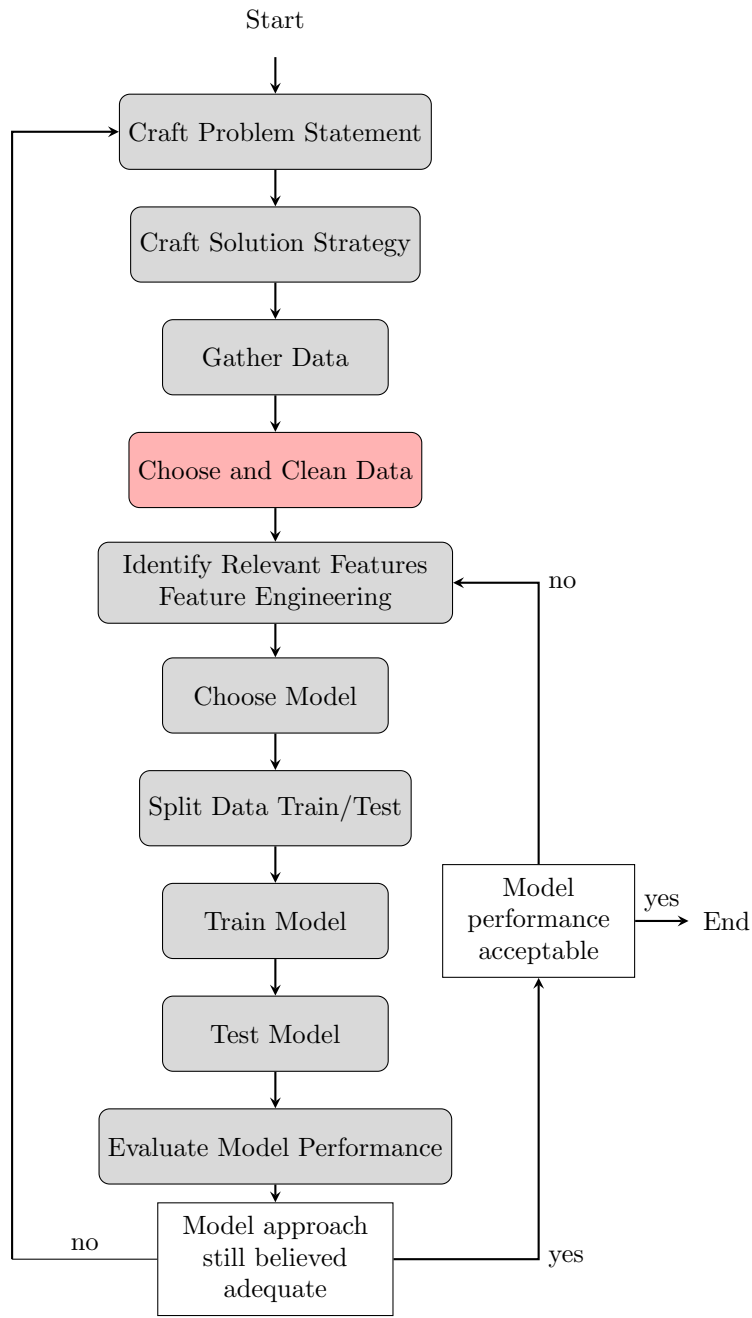

Next we consider the "Choose and Clean Data" stage in the ML Process Flow (see figure \ref{fig:workflow3}). 

Raw data is rarely in the format needed for a machine learning algorithm to consume. Choosing and cleaning up the data may involve identifying features, converting data into integer or Boolean representation, identifying missing data, or removing uninteresting data. 

A feature is an input value that describes characteristics of labels in the dataset.  A feature is an input (often a field of a spreadsheet or a form) and a label is an output. Two examples of features are categorical and numerical. For example, when working with software defect data, some fields in the dataset might include: title, description, severity, test phase, test type, creator, owner, date, and environment. The severity might contain numeric values (1, 2, and 3, for example), while the environment field may contain categorical values (dev, test, and prod, for example). One example of creating a feature that is different than just the dataset's fields is determining if the defect record was created by a person or automated process. For example, if all automated IDs had a detectable pattern in their name (such as auto\_id@company.com), the creator field can be used to create a new feature called ("human\_creator") with Boolean values.

If you do not have expert knowledge of the data that is being used to train a machine learning algorithm, then you will want to consult with an SME to gain a better understanding of the dataset being used.  This activity is often part of data auditing and analysis.  Following are a set of questions that you can ask the SME to help you better understand the dataset and thus help you clean the data so that it can be consumed by a machine learning algorithm.
\begin{enumerate}
    \item Are there fields in the data that are unreliable? (eg. human entered data)
    \item Are there fields in the data that are redundant? (eg. it may be possible to reduce fields for a defect record creator, their email, and their employee ID down to one)
    \item Are there fields in the data that are misleading? (eg. data that has been edited after the fact)
    \item Are there fields in the data that are biased? (eg. a field that only has data when opened by a particular team)
    \item Are there fields in the data that can be converted and/or combined to new features that provide more (or different) values? (eg. an individual date field that can transform into a month or quarter)
    \item Are there values that can be modified into something that not only increases the machine learning model performance, but improves explain-ability?  (eg. a specific code level that can be grouped into either a 'dev' or 'prod' category)
\end{enumerate}

\begin{example} \label{DEV-GA}
    Suppose we have a field in our dataset with the following values:
    \begin{itemize}
        \item A-DEV-123
        \item A-DEV-456
        \item B-DEV-890
        \item A-GA-1
        \item B-GA-1
        \item B-GA-3
    \end{itemize}

    After talking with the SME, we learn that any value that has DEV in it represents data that was collected while the software was in the development cycle (before it was released to the customer), and any value that has GA (General Availability) in it represents data that was collected after the software was released to the customer. We can apply expertise to this dataset and collapse these values into two possibilities: DEV and GA, representing the time that the data was collected. By transforming the original data into a new, binary feature, we hope to reduce noise and increase the number of examples.
    
\end{example}

Consulting an SME is not the only way to gain understanding of the fields in the dataset. We can also understand the data by applying simple math and statistics to the data. This step of the data auditing process may best be done visually, by looping over every field and charting the data. For each field, we can examine the shape of the distribution, the outliers, the singletons, and the values that are most or least frequent. When doing this, we can also attempt to display the correlation of different features and determine if certain features are highly related to others and should be addressed (covered below in figure \ref{code:create_heatmap}).

Consider the table (table \ref{tab:dataset_fields}) and histogram (figure \ref{hist:dataset_fields}) below. The table is the tabular representation of the fields that make up the dataset. It is often helpful to view the data this way, but it may be difficult to determine how data for a particular field is distributed.

\begin{center}
\begin{table}[H]
    \begin{tabular}{||c c c c c||} 
     \hline
      & Defect & Severity & Test\_Type & Test\_Phase \\ [0.5ex] 
     \hline\hline
     0 & \#001 & 4 & Regression & Acceptance \\ 
     \hline
     1 & \#002 & 3 & Regression & Function \\ 
     \hline
     2 & \#003 & 4 & Stress & Function \\
     \hline
     3 & \#004 & 1 & Regression & Unit \\
     \hline
     4 & \#005 & 2 & Recovery & Solution \\
     \hline
     5 & \#006 & 4 & Stress & Acceptance \\
     \hline
     6 & \#007 & 3 & Stress & System \\
     \hline
     7 & \#008 & 1 & Recovery & System \\
     \hline
     8 & \#009 & 3 & Regression & Solution \\
     \hline
     9 & \#010 & 2 & Regression & Function \\ [1ex] 
     \hline
    \end{tabular}
    \caption{Example of what tabular data might look like.}
    \label{tab:dataset_fields}
\end{table}
\end{center}

If we wanted to understand how often different Test Phases are seen in our dataset, we could display the data as a histogram (figure \ref{hist:dataset_fields}). This visual representation helps contextualize the frequency counts of each Test Phase with respect to the other values.

\pgfplotstableread[row sep=\\,col sep=&]{
    interval & feat \\
    Unit   & 1 \\
    Function   & 3 \\
    System   & 2 \\
    Solution   & 1 \\
    Acceptance   & 3 \\
    }\featimp
\pgfplotsset{xtick style={draw=none}}
\begin{figure}[H]
\begin{tikzpicture}
    \begin{axis}[
            ybar,
            symbolic x coords={Unit, Function, System, Solution, Acceptance},
            xtick=data,
            xticklabels={Unit, Function, System, Solution,  Accept}
        ]
        \addplot table[x=interval,y=feat]{\featimp};
    \end{axis}
\end{tikzpicture}
\caption{The distribution of values from the Test Phase field.}
\label{hist:dataset_fields}
\end{figure}
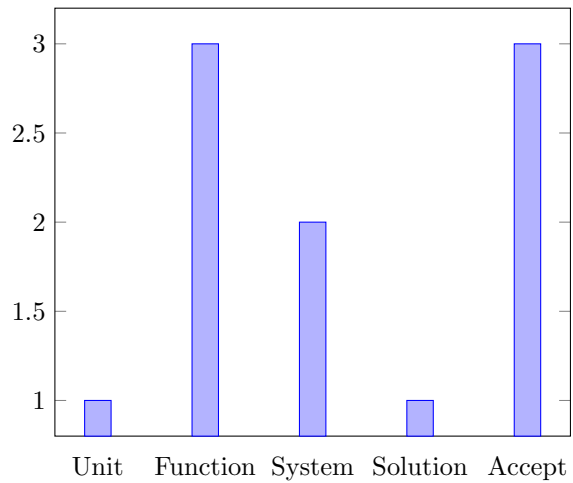

More questions come up as we started to better understand the dataset, and in the next few sections, we will provide additional details for addressing some of them.

\begin{itemize}
    \item What do we do with missing data?
    \begin{itemize}
        \item Is it just a few rows of data that is missing?
        \item Does most of the data omit a certain field?
        \item Can we try different types of imputation (which is the process of filling in mising data)?
        \item If we will try imputation, could that introduce bias?
    \end{itemize}
    \item What do we do with "bad data" that is incorrect or inconsistent?
    \begin{itemize}
        \item Is the bad data obvious to spot?
        \item How many rows contain bad data?
        \item Is the whole field at risk of being unreliable?
        \item Could this field turn into a different one?
    \end{itemize}
    \item How do we handle the outliers?
    \begin{itemize}
        \item Should we remove the outliers?
        \item Should we keep the outliers for context?
        \item Could outliers affect the model in a way in which they should be removed?
    \end{itemize}
    \item Is this field valuable?
    \begin{itemize}
        \item Are all the values for a particular field the same?
        \item Could this field, with additional information, be transformed into a more useful feature?
    \end{itemize}
\end{itemize}

It is important to transform fields whose values cannot be easily consumed by the machine learning model. Some models cannot interpret text, but if you convert certain fields to numeric ones, then there may be assumed ordinality. Below we will describe two example fields related to software defect data that we converted from categorical to numeric values. In the \hyperref[TestPhaseEncoding]{Test Phase} example, we elected to convert the phase values to incrementing integers since software often goes through different test phases in a defined order. However, in the \hyperref[TestTypeExample]{Test Type} example, we needed to take additional steps to remove any assumed ordinality.

\begin{example} \label{TestPhaseEncoding}
  Suppose we have a field that describes the phase of test a defect was found in. Values for for this field may include: UNIT, FUNCTION, SYSTEM, SOLUTION, and ACCEPTANCE. The machine learning model cannot handle text values, so we transform them to be 1, 2, 3, 4, 5. Since Unit Test comes before Function Test, and so on, we can use the converted numbers as-is. If a machine learning model checks to see if the test phase is $>2$, this would correlate to asking if the reported problem was found in System Test or later.
\end{example}

\begin{example} \label{TestTypeExample}
  Suppose we have a field that describes the type of test that discovered a defect. Values for for this field may include: RECOVERY, STRESS, and REGRESSION. The machine learning model cannot handle text values, so we transform them to be 1, 2, 3. Recovery tests are not "less" than stress tests, but a model might try to create a decision on if the test type is $<2$. In this case, one way to engineer this feature is to \hyperref[subsec:encoding]{OneHot Encode} it and turn one field of test types into three fields. RECOVERY, STRESS, and REGRESSION will each be their own feature and will have Boolean values of 0 or 1 to represent false or true for the test type.
\end{example}

While converting categorical fields to numeric ones is important, it is still necessary to review the numeric fields. We may find that some data needs to be standardized. From \href{https://scikit-learn.org/stable/modules/generated/sklearn.preprocessing.StandardScaler.html}{sklearn's website}: "Standardization of a dataset is a common requirement for many machine learning estimators: they might behave badly if the individual features do not more or less look like standard normally distributed data (e.g. Gaussian with 0 mean and unit variance)." Since some models make assumptions about the data, standardization may be required to ensure the data matches those expectations.

The following list is a summary of our lessons learned when cleaning up the data.

\begin{itemize}
    \item Data cleansing is a requirement for any machine learning project.
    \item Cleaning the data is where most of the work happens with machine learning.
    \item Data scientist may not have the understanding of the data needed to clean it appropriately so they should consult an SME.
    \item SME can show logical groupings, correlation, and patterns that may not be obvious when looking at the data in isolation.
    \item SME may be able to provide outside knowledge that can be applied to the data.
    \item There are not always clear answers for cleansing the data.
    \item Much of the cleaning of the data is an iterative set of experiments to engineer the features that work with a particular dataset and machine learning model.
    \item Simple questions tend to turn into hard answers of "It depends" and then more analysis of the data is required to drill down into the actual answer of the question.
    \item Domain knowledge applied to data can transform a data feature from being not very important to an important feature.
\end{itemize}

Feature engineering, which is the process of transforming or adding features, can be used to normalize the input into the machine learning model, help clean up the existing dataset (examples in both example \ref{DEV-GA} and figure \ref{TestPhaseEncoding}), and turn a valueless field into an important feature. It is even possible to combine multiple fields together to create a single new feature. In the next section, we will go over how feature engineering can be performed to add features to the dataset based on additional information.

\newpage
\subsection{What to do with highly correlated data}

\begin{figure} [H]
  \begin{tikzpicture} [node distance=1.5cm]
    \node(Start) [BegEnd] {Start};
    \node(PS) [startstop, below of=Start] {Craft Problem Statement};
    \node(CS) [startstop, below of=PS] {Craft Solution Strategy};
    \node(GatherData) [startstop, below of=CS] {Gather Data};
    \node(ChooseData) [startstop, below of=GatherData] {Choose and Clean Data};
    \node(Feature) [focus, below of=ChooseData]{
      \begin{tabular}{c}
      Identify Relevant Features  \\
      Feature Engineering 
      \end{tabular}
      };
    \node(ChooseModel) [startstop, below of=Feature] {Choose Model};
    \node(SplitData) [startstop, below of=ChooseModel] {Split Data Train/Test};
    \node(TrainModel) [startstop, below of=SplitData] {Train Model};

    \node(TestModel) [startstop, below of=TrainModel] {Test Model};
    \node(EP) [startstop, below of=TestModel] {Evaluate Model Performance};
    \node(DMG) [draw, below of=EP]{
      \begin{tabular}{c}
      Model approach  \\
      still believed \\
      adequate
      \end{tabular}
      };
    \node(DMF) [draw, right of=TrainModel, xshift=2cm]{
      \begin{tabular}{c}
      Model \\
      performance \\
      acceptable
      \end{tabular}
      };
    \node(End) [BegEnd, right of=DMF,xshift=1cm] {End};  
    \node(blank) [BegEnd, left of=DMG, xshift=-2cm] { };
    \draw [arrow] (Start) -- (PS);
    \draw [arrow] (PS) -- (CS);
    \draw [arrow] (CS) -- (GatherData);
    \draw [arrow] (GatherData) -- (ChooseData);
    \draw [arrow] (ChooseData) -- (Feature);
    \draw [arrow] (Feature) -- (ChooseModel);
    \draw [arrow] (ChooseModel) -- (SplitData);
    \draw [arrow] (SplitData) -- (TrainModel);
    \draw [arrow] (TrainModel) -- (TestModel);
    \draw [arrow] (TestModel) -- (EP);
    \draw [arrow] (EP) -- (DMG);
    \draw [-] (DMG) to node[anchor=south] {no} (blank.center);
    \draw [arrow] (blank.center) |- (PS);
    \draw [arrow] (DMF) |- node[anchor=west] {no} (Feature);
    \draw [arrow] (DMG) -| node[anchor=west] {yes} (DMF);
    \draw [arrow] (DMF) -- node[anchor=south] {yes} (End);
  \end{tikzpicture}
  \caption{Feature Engineering}
  \label{fig:workflow4}
\end{figure}
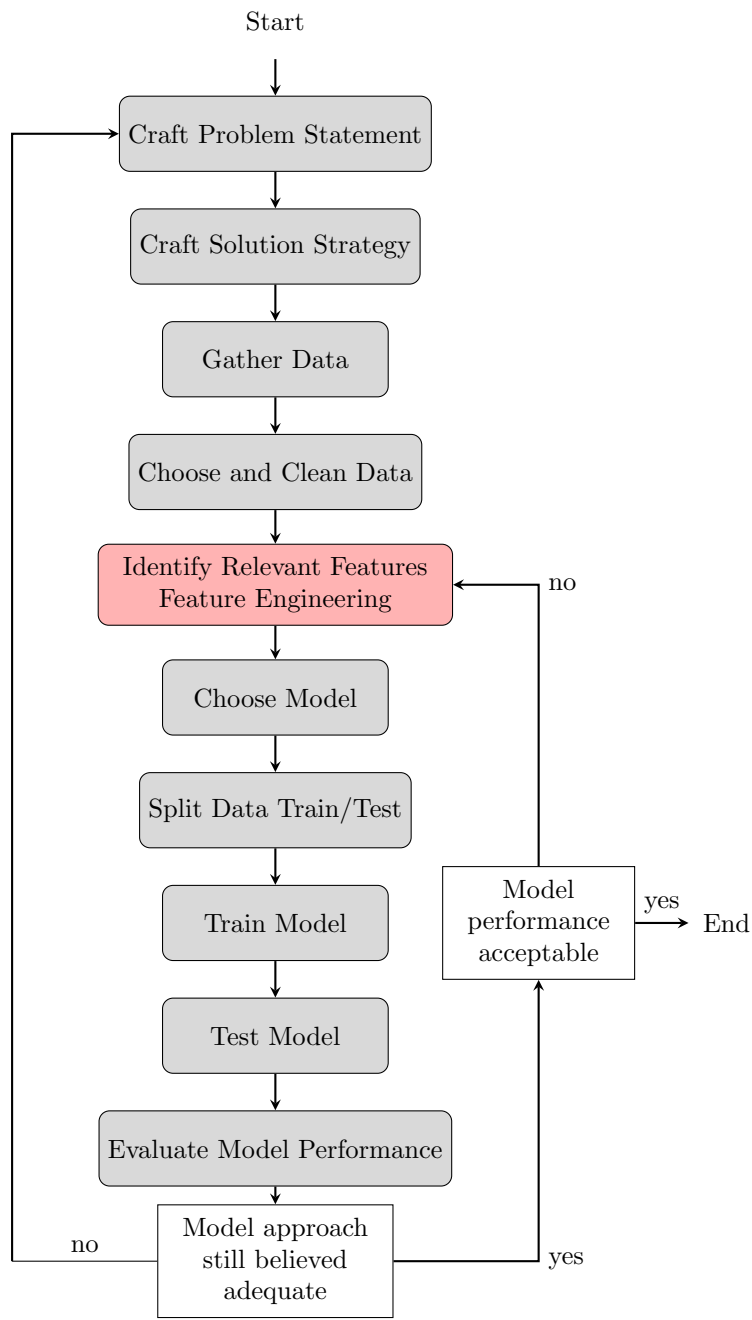

With data selected and cleaned, we can move on to the  "Identify Relevant Features / Feature Engineering" stage in the ML Process Flow (see figure \ref{fig:workflow4}). 

One important part of featuring engineering is examining the data to identify highly correlated features. If multiple features are highly correlated, problems can arise. These problems can range from simple redundancy to inaccurate model behavior and poor performance. In the case of regression models, highly correlated features can lead to multicollinearity, a problem in which it can be difficult to identify the impact of individual variables.

\begin{figure}[h!]
    \centering
    \includegraphics[width=6cm]{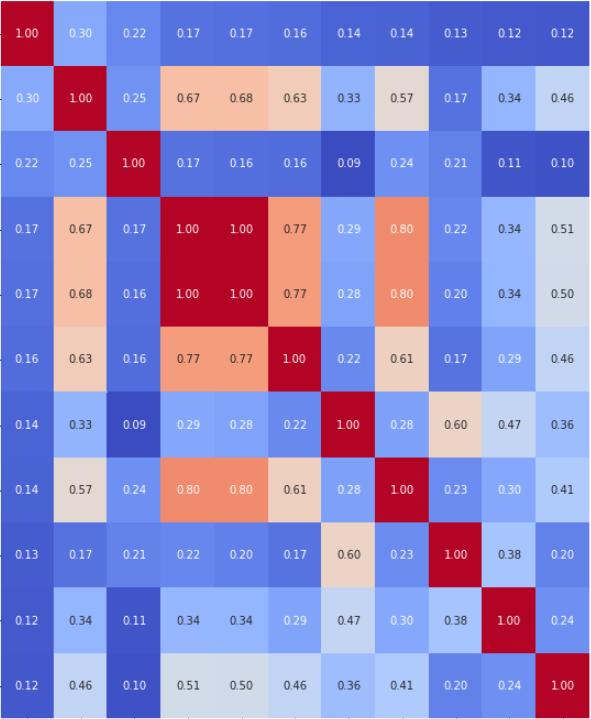}
    \caption{A heatmap showing feature correlation.}
    \label{fig:heatmap.png}
\end{figure}

There are multiple techniques and approaches for determining correlation. The \href{https://pandas.pydata.org/}{pandas} Python package has a `corr()` function for computing pairwise correlations using various methods. While the data can be plotted to help visually depict correlation, we have found it valuable to display the data as a matrix. Creating a heatmap (as seen in figure \ref{fig:heatmap.png}) is an easy way to quickly identify correlated features as well as how highly correlated they are.

Assuming you have a pandas dataframe called df, the below code example (in figure \ref{code:create_heatmap}) may be used to create a heatmap of feature correlations.

\begin{figure} [H]
\begin{lstlisting}[language=Python]
import matplotlib.pyplot as plt
import seaborn as sns

# Heatmap of correlations
plt.subplots(figsize=(12, 12))
g = sns.heatmap(df.corr(), annot=True, fmt = ".2f", vmin=-1, vmax=1)
\end{lstlisting}
\caption{Example of using a heatmap to display feature correlations}
\label{code:create_heatmap}
\end{figure}

Running that snippet will result in a visual representation of how strongly each feature correlates to the others. In the example output below (in figure \ref{fig:heatmap.png}), we see different shades of blue for the low correlations, but oranges and reds for features that are strongly correlated.

When examining the data for correlations, it is important to think through the features, their meanings, and why correlation may exist. Just because features are correlated does not mean that there exists causation. If an input feature has correlation with the target feature, it may be an important feature for a model's prediction. However, if multiple input features are correlated with each other, it may be beneficial to only provide one to a model in order to avoid bias.

\subsection{Identifying missing data}
\label{subsec:identifying-missing-data}
One important step in data preparation is the handling of missing data, and this was especially true for us since we planned on using the algorithms provided by \href{https://scikit-learn.org/}{scikit-learn}. Model performance will likely be impacted when data is missing, and there are a few ways to handle this. Five common methods are:
\begin{enumerate}
    \item Remove all records that are missing any value. If there aren't many records with missing values, this can be an easy solution.
    \item Remove features that are commonly missing from the data. If almost all records are missing values for a feature, it may be best to remove the entire feature.
    \item Impute the value for the missing data. This is only practical when there are only a few records with the missing values for a given feature. Though there are various methods for imputation, it may be possible to calculate or predict a value for the missing data. 
    \item Create a new category or categories such as "Missing" or "Filled" to replace all missing or invalid data.
    \item Do nothing and leave a field with missing data.
\end{enumerate}

In addition to missing data, it is possible that some values in the dataset contain false, incorrect, inconsistent, or otherwise nonsensical data. While it can be difficult to identify without the help of an SME, it is important to address this problem. Unusable data can negatively impact a model's performance just as much, if not more, than having missing data in the dataset. If incorrect data cannot be transformed, it may be beneficial to remove the unusable value and proceed as if the data was missing altogether.

\subsection{Encoding}
\label{subsec:encoding}

While most machine learning models can handle Boolean values and numbers, many cannot take categorical values (such as strings of text) as input. Therefore, it is important to transform values into numerical ones. This process is called encoding. There are different ways to encode, and scikit-learn's \href{https://scikit-learn.org/stable/modules/classes.html#module-sklearn.preprocessing}{preprocessing} module has many options. One common method of encoding (and one that we used in our projects) is called OneHot Encoding.

OneHot Encoding is the process of turning a single feature of categorical values into N numerical values, where N is the number of unique original values. The end result is an additional feature for each value that contains only 0s and 1s. This new form is more appropriate for a model to handle and can increase the performance.

Below is an example of how we used OneHot Encoding when working with software defect data. To start, here is a sample of code that displays a dataframe before OneHot Encoding the data.
\begin{figure} [H]
\begin{lstlisting}[language=Python]
import pandas as pd

example_df = pd.DataFrame.from_dict(
    {
        "Defect": ["#001", "#002", "#003", "#004", "#005",
                   "#006", "#007", "#008", "#009", "#010"],
        "Severity": [4, 3, 4, 1, 2, 4, 3, 1, 3, 2],
        "Area": ["Memory", "I/O", "Network", "Network",
                 "Memory", "Filesystem", "Threading",
                 "I/O", "Threading", "Memory"]
    }
)
example_df
\end{lstlisting}
\caption{Example defect data before OneHot Encoding}
\label{code:ExpBeforeOneHot}
\end{figure}

Running the above code in figure \ref{code:ExpBeforeOneHot} displays our dataframe, and we can see in the table below that the data for the Area field contains various different values of text data.

\begin{center}
\begin{table}[H]
    \begin{tabular}{||c c c c||} 
     \hline
      & Defect & Severity & Area \\ [0.5ex] 
     \hline\hline
     0 & \#001 & 4 & Memory \\ 
     \hline
     1 & \#002 & 3 & I/O \\ 
     \hline
     2 & \#003 & 4 & Network \\
     \hline
     3 & \#004 & 1 & Network \\
     \hline
     4 & \#005 & 2 & Memory \\
     \hline
     5 & \#006 & 4 & Filesystem \\
     \hline
     6 & \#007 & 3 & Threading \\
     \hline
     7 & \#008 & 1 & I/O \\
     \hline
     8 & \#009 & 3 & Threading \\
     \hline
     9 & \#010 & 2 & Memory \\ [1ex] 
     \hline
    \end{tabular}
    \caption{A Pandas Dataframe showing example data, including an Area field.}
\end{table}
\end{center}

To OneHot Encode the the Area field, we use the following code snippet. This will create five new features (Memory, I/O, Network, Filesystem, and Threading) that will replace the original Area feature.

\begin{figure} [H]
\begin{lstlisting}[language=Python]
to_onehot_encode = ["Area"]
pd.get_dummies(example_df, prefix=to_onehot_encode, columns=to_onehot_encode)
\end{lstlisting}
\caption{Example defect data after OneHot Encoding}
\label{code:ExpAfterOneHot}
\end{figure}

 When we run our second code sample in figure \ref{code:ExpAfterOneHot}, we see that the data for the Area field has been expanded into numerous different fields with a binary "yes or no" value, represented in the output below as 1s and 0s.

\begin{center}
\begin{table}[H]
    \begin{tabular}{||c c c c c c c c||} 
     \hline
      & Defect & Severity & Area\_Filesystem & Area\_I/O & Area\_Memory & Area\_Network & Area\_Threading \\ [0.5ex] 
     \hline\hline
     0 & \#001 & 4 & 0 & 0 & 1 & 0 & 0 \\ 
     \hline
     1 & \#002 & 3 & 0 & 1 & 0 & 0 & 0 \\ 
     \hline
     2 & \#003 & 4 & 0 & 0 & 0 & 1 & 0 \\ 
     \hline
     3 & \#004 & 1 & 0 & 0 & 0 & 1 & 0 \\ 
     \hline
     4 & \#005 & 2 & 0 & 0 & 1 & 0 & 0 \\ 
     \hline
     5 & \#006 & 4 & 1 & 0 & 0 & 0 & 0 \\ 
     \hline
     6 & \#007 & 3 & 0 & 0 & 0 & 0 & 1 \\ 
     \hline
     7 & \#008 & 1 & 0 & 1 & 0 & 0 & 0 \\ 
     \hline
     8 & \#009 & 3 & 0 & 0 & 0 & 0 & 1 \\ 
     \hline
     9 & \#010 & 2 & 0 & 0 & 1 & 0 & 0 \\ [1ex] 
     \hline
    \end{tabular}
\caption{A Pandas Dataframe showing that the Area column has been replaced with OneHot Encoded values.}
\end{table}
\end{center}

\subsection{Applicability to other areas}
One project we worked on attempted to predict service (software updates) that would introduce an error, with the hope that the identified items could be reviewed, tested, and given more attention than normal before being made available to customers. The project consumed data from a common database that held service information across multiple brands, products and teams, but to start, we focused just on the Project 1 data for initial feasibility analysis. The initial findings showed that the project was worth pursuing further, and in an attempt to improve the model's performance, we brought in other Project 1-specific data sources. The data from these new sources quickly became some of the most important features for the model's predictions, and yielded the improvement we were hoping for.

As we began discussions with other teams outside of Project 1, we saw interest in applying our techniques to their data. However, as conversations about data unfolded, we realized that different products and platforms collected different data during the development process than others. The data sources that gave the model its most important features did not exist outside of the Project 1 development team, and other areas struggled to identify if they even could collect certain metrics. A few areas realized that they could start collecting more data, but wanted to first see how the model would eventually pan out in Project 1 before they undertook the effort of changing their processes. We realized that even if these other development teams began collecting data immediately, it could be months or years before there was enough of a sample to be usable, and even then we could not be sure that machine learning could accurately make predictions on their data. Though obvious in hindsight, this was an important lesson that even though an idea may be applicable across different products and platforms, teams operate differently and record different metrics and statistics. It is all about the data...

\newpage
\section{Modeling Choices}

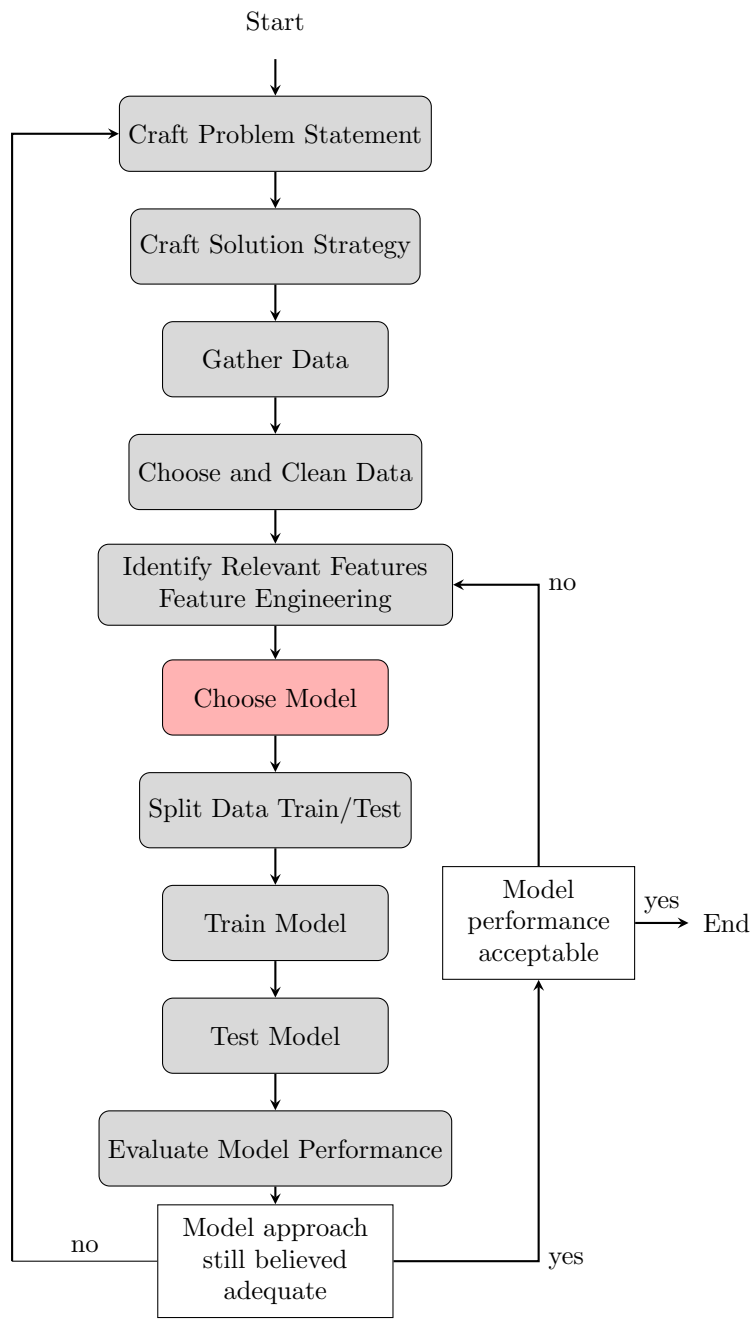
\begin{figure} [H]
  \begin{tikzpicture} [node distance=1.5cm]
    \node(Start) [BegEnd] {Start};
    \node(PS) [startstop, below of=Start] {Craft Problem Statement};
    \node(CS) [startstop, below of=PS] {Craft Solution Strategy};
    \node(GatherData) [startstop, below of=CS] {Gather Data};
    \node(ChooseData) [startstop, below of=GatherData] {Choose and Clean Data};
    \node(Feature) [startstop, below of=ChooseData]{
      \begin{tabular}{c}
      Identify Relevant Features  \\
      Feature Engineering 
      \end{tabular}
      };
    \node(ChooseModel) [focus, below of=Feature] {Choose Model};
    \node(SplitData) [startstop, below of=ChooseModel] {Split Data Train/Test};
    \node(TrainModel) [startstop, below of=SplitData] {Train Model};

    \node(TestModel) [startstop, below of=TrainModel] {Test Model};
    \node(EP) [startstop, below of=TestModel] {Evaluate Model Performance};
    \node(DMG) [draw, below of=EP]{
      \begin{tabular}{c}
      Model approach  \\
      still believed \\
      adequate
      \end{tabular}
      };
    \node(DMF) [draw, right of=TrainModel, xshift=2cm]{
      \begin{tabular}{c}
      Model \\
      performance \\
      acceptable
      \end{tabular}
      };
    \node(End) [BegEnd, right of=DMF,xshift=1cm] {End};  
    \node(blank) [BegEnd, left of=DMG, xshift=-2cm] { };
    \draw [arrow] (Start) -- (PS);
    \draw [arrow] (PS) -- (CS);
    \draw [arrow] (CS) -- (GatherData);
    \draw [arrow] (GatherData) -- (ChooseData);
    \draw [arrow] (ChooseData) -- (Feature);
    \draw [arrow] (Feature) -- (ChooseModel);
    \draw [arrow] (ChooseModel) -- (SplitData);
    \draw [arrow] (SplitData) -- (TrainModel);
    \draw [arrow] (TrainModel) -- (TestModel);
    \draw [arrow] (TestModel) -- (EP);
    \draw [arrow] (EP) -- (DMG);
    \draw [-] (DMG) to node[anchor=south] {no} (blank.center);
    \draw [arrow] (blank.center) |- (PS);
    \draw [arrow] (DMF) |- node[anchor=west] {no} (Feature);
    \draw [arrow] (DMG) -| node[anchor=west] {yes} (DMF);
    \draw [arrow] (DMF) -- node[anchor=south] {yes} (End);
  \end{tikzpicture}
  \caption{Choose Model}
  \label{fig:workflow5}
\end{figure}

With the data and features prepared, we move on to the "Choose Model" stage in the ML Process flow (see figure \ref{fig:workflow5}). 

When addressing a machine learning problem, it is important to consider what algorithms are appropriate for the task. There is a \href{https://scikit-learn.org/stable/tutorial/machine_learning_map/index.html}{decision tree} on the scikit-learn website about choosing the most appropriate option.

\subsection{Classification vs. Regression}

One simple way of understanding the difference between classification and regression problems is to think about the prediction type of the model. Are we predicting a numeric value or a percentage, such as the percentage of users that are likely to be impacted by a defect? If so, then this is likely a regression problem. Are we predicting a label, or class, such as if a defect is a duplicate or not? If so, then this is likely a classification problem.

Throughout this paper, we have referenced two problems which we sought machine learning solutions for, and both cases are classification problems. In the example of predicting fixes that contain errors, we were trying to predict if the fix falls in the "problem" class or "no problem" class. In the example of predicting problems that are duplicates, we were trying to predict if the problem falls in the "duplicate" class or "not duplicate" class. These are both cases of trying to predict a single, binary variable. It is possible to build a classifier that predicts multiple classes. While we did not have a need to explore these, scikit-learn does provide information on their \href{https://scikit-learn.org/stable/modules/multiclass.html}{"Multiclass and multioutput algorithms"} page.

\subsection{Picking an Initial Model}

\begin{figure}[h!]
    \centering
    \includegraphics[width=6cm]{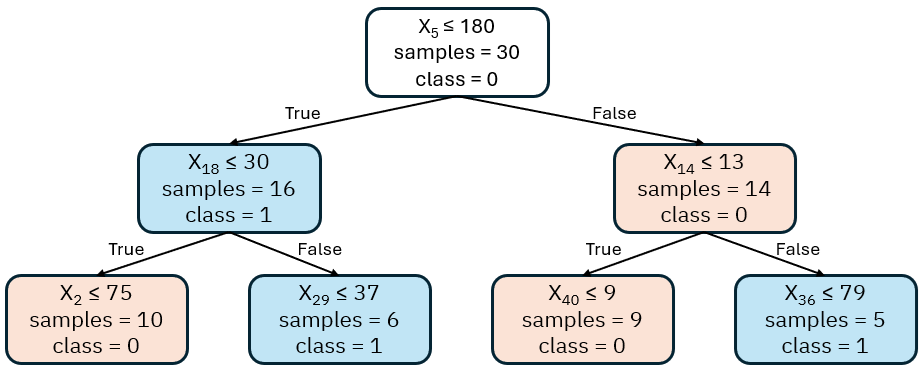}
    \caption{A visual representation of a decision tree.}
    \label{fig:decision_tree.png}
\end{figure}

Since our goal was to predict a single, binary variable, we elected to start with a Decision Tree Classifier since it was so approachable. Instead of an enigmatic algorithm, a Decision Tree Classifier has high explainability and there are many visual ways of looking at how a model works as seen in figure \ref{fig:decision_tree.png}. This was very important for us since we were new to the process, so having some visual helped us learn what the algorithm was creating and how it was deciding.

\subsection{Training a Model}

\begin{figure} [H]
  \begin{tikzpicture} [node distance=1.5cm]
    \node(Start) [BegEnd] {Start};
    \node(PS) [startstop, below of=Start] {Craft Problem Statement};
    \node(CS) [startstop, below of=PS] {Craft Solution Strategy};
    \node(GatherData) [startstop, below of=CS] {Gather Data};
    \node(ChooseData) [startstop, below of=GatherData] {Choose and Clean Data};
    \node(Feature) [startstop, below of=ChooseData]{
      \begin{tabular}{c}
      Identify Relevant Features  \\
      Feature Engineering 
      \end{tabular}
      };
    \node(ChooseModel) [startstop, below of=Feature] {Choose Model};
    \node(SplitData) [focus, below of=ChooseModel] {Split Data Train/Test};
    \node(TrainModel) [focus, below of=SplitData] {Train Model};

    \node(TestModel) [startstop, below of=TrainModel] {Test Model};
    \node(EP) [startstop, below of=TestModel] {Evaluate Model Performance};
    \node(DMG) [draw, below of=EP]{
      \begin{tabular}{c}
      Model approach  \\
      still believed \\
      adequate
      \end{tabular}
      };
    \node(DMF) [draw, right of=TrainModel, xshift=2cm]{
      \begin{tabular}{c}
      Model \\
      performance \\
      acceptable
      \end{tabular}
      };
    \node(End) [BegEnd, right of=DMF,xshift=1cm] {End};  
    \node(blank) [BegEnd, left of=DMG, xshift=-2cm] { };
    \draw [arrow] (Start) -- (PS);
    \draw [arrow] (PS) -- (CS);
    \draw [arrow] (CS) -- (GatherData);
    \draw [arrow] (GatherData) -- (ChooseData);
    \draw [arrow] (ChooseData) -- (Feature);
    \draw [arrow] (Feature) -- (ChooseModel);
    \draw [arrow] (ChooseModel) -- (SplitData);
    \draw [arrow] (SplitData) -- (TrainModel);
    \draw [arrow] (TrainModel) -- (TestModel);
    \draw [arrow] (TestModel) -- (EP);
    \draw [arrow] (EP) -- (DMG);
    \draw [-] (DMG) to node[anchor=south] {no} (blank.center);
    \draw [arrow] (blank.center) |- (PS);
    \draw [arrow] (DMF) |- node[anchor=west] {no} (Feature);
    \draw [arrow] (DMG) -| node[anchor=west] {yes} (DMF);
    \draw [arrow] (DMF) -- node[anchor=south] {yes} (End);
  \end{tikzpicture}
  \caption{Train Model}
  \label{fig:workflow6}
\end{figure}
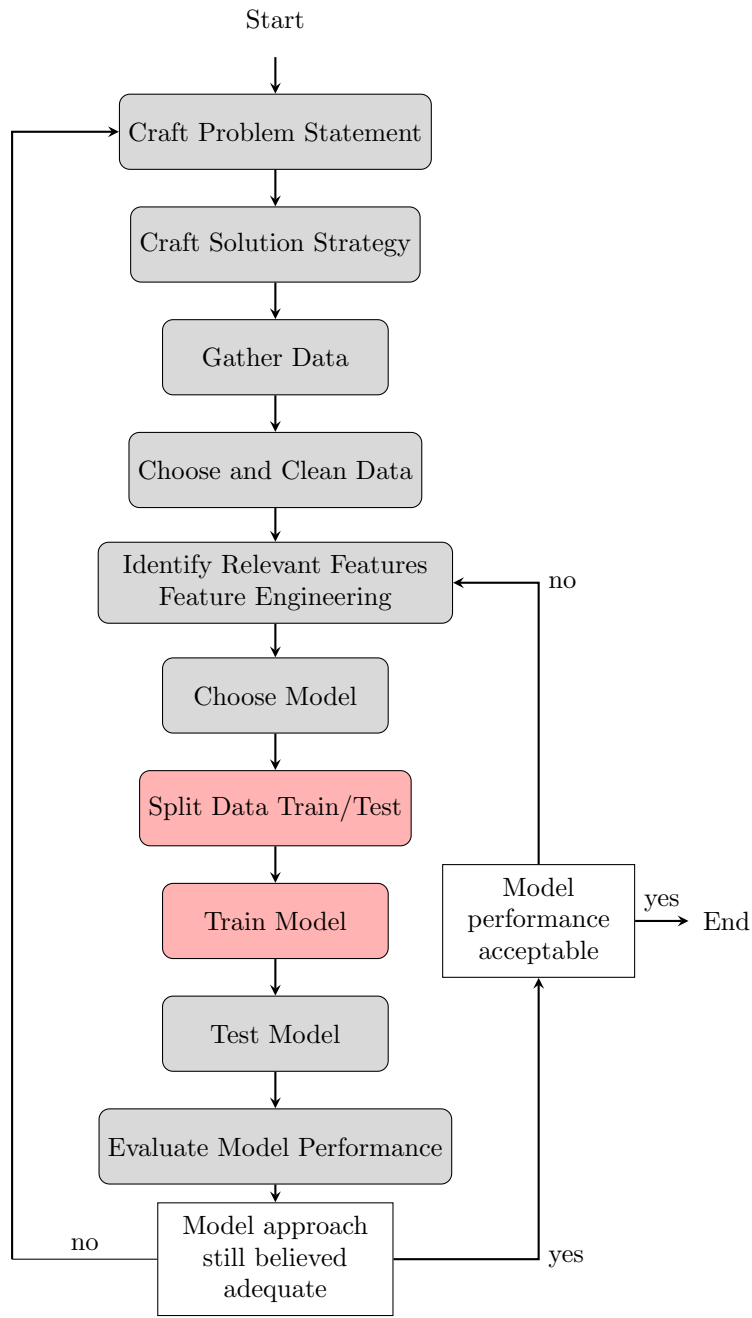

After a model choice has been made, we moved on to the "Split Data Train/Test" and "Train Model" stages in the ML Process Flow (see figure \ref{fig:workflow6}). 

Our first step was to split the data into two separate datasets: one for training and one for testing. The general, accepted rule-of-thumb is that 80\% of the data should be used for training the model while the other 20\% should be used for testing the model's performance. Since we were using the Python packages from scikit-learn, we were able to easily split our dataset using the sklearn.model\textunderscore selection.train\textunderscore test\textunderscore split function, as seen in the code example below in figure \ref{code:traintestsplit}. In the code example, we used default settings, but you may need to explore adjusting the defaults for different criteria.

\begin{figure}[H]
\begin{lstlisting}[language=Python]
from sklearn.model_selection import train_test_split

# Independent variables
X = example_df.drop('To_Predict_On', axis=1)
# Dependent variable
y = example_df['To_Predict_On']  

X_train, X_test, y_train, y_test = train_test_split(X, y, test_size=0.2, random_state=1)
\end{lstlisting}
\caption{Splitting data into training and testing datasets.}
\label{code:traintestsplit}
\end{figure}

After splitting our data into training and testing datasets, we built our classifier using the training set. In the code example below in figure \ref{code:buildDTC}, we use a Decision Tree for our classifier.

\begin{figure}[H]
\begin{lstlisting}[language=Python]
from sklearn.tree import DecisionTreeClassifier

# Create and train a Decision Tree Classifier
dt_classifier = DecisionTreeClassifier()
dt_classifier.fit(X_train, y_train)
\end{lstlisting}
\caption{Building a Decision Tree Classifier.}
\label{code:buildDTC}
\end{figure}

At this point, we could either begin making predictions against the testing dataset to see how we perform, or we could look deeper into the model we built.

Starting with the latter, the first thing we did was display the feature importance metrics for our model. Each feature was given a value of how impactful that feature was to the model's overall decision making, with a higher number indicating more importance. This data showed us how meaningful certain features were for the model.

The code snippet shown below in figure \ref{code:DisplayFeatureImportance} serves as an example way to graph the feature important metrics of a model.

\begin{figure} [H]
\begin{lstlisting}[language=Python]
import matplotlib.pyplot as plt
import pandas as pd

# Feature importances for the Decision Tree Classifier
# Variable fi is the feature importance metrics
fi = pd.Series(dt_classifier.feature_importances_,
               index=X_train.columns).sort_values(ascending=False)
fi.head(10).plot(kind='bar', title="Top 10 important features")
\end{lstlisting}
\caption{Displaying feature importance metrics of the Decision Tree Classifier}
\label{code:DisplayFeatureImportance}
\end{figure}

Graphing the feature importance data is an easy way to visually understand the data, as shown in the resulting figure \ref{code:FeatureImportanceGraph}. It is important to review which values are important to the model and consider what that means. Do these features align with the expectations of SMEs? Are there new insights to learn from this data? Does the model need to be adjusted? Are certain features inherently biased?

\pgfplotstableread[row sep=\\,col sep=&]{
    interval & feat \\
    Feat1   & .48 \\
    Feat2   & .28 \\
    Feat3   & .17 \\
    Feat4   & .10 \\
    Feat5   & .08 \\
    Feat6   & .02 \\
    Feat7   & .01 \\
    Feat8   & .008 \\
    Feat9   & .006 \\
    Feat10  & .003 \\
    }\featimp
\pgfplotsset{xtick style={draw=none}}
\begin{figure}[H]
\begin{tikzpicture}
    \begin{axis}[
            ybar,
            symbolic x coords={Feat1, Feat2, Feat3, Feat4, Feat5, Feat6, Feat7, Feat8, Feat9, Feat10},
            xtick=data,
            xticklabels={,,}
        ]
        \addplot table[x=interval,y=feat]{\featimp};
    \end{axis}
\end{tikzpicture}
\caption{A graph showing feature importance values from a Decision Tree Classifier.}
\label{code:FeatureImportanceGraph}
\end{figure}
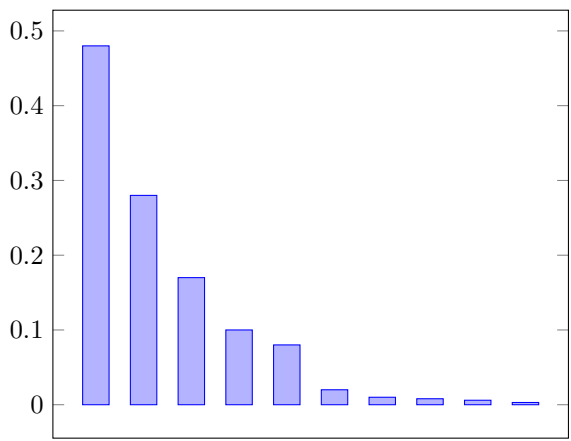

Additionally, we can construct a visual representation of the classifier using the nodes and edges. This can be useful as a means of manually exploring the branches and decisions that will be used when predicting on new data. We found it important to analyze the tree to see if the features and values used for the decisions aligned with domain expertise. However, we also looked for things that did not align with our expectations, with the goal of either gaining new insights or finding errors with the model's algorithm.

After exploring the feature importance metrics and the tree visualization, we decided to test our model's performance by having it make predictions against the testing dataset. Since we knew the target labels for this data, we could compare these values against the predicted values to see how well the model performed. In the next section, we will describe two ways that we used to check our model's results: \hyperref[subsec:confusionmatrices]{Confusion Matrices and Classification Reports}.

\subsection{Testing the Model}

\begin{figure} [H]
  \begin{tikzpicture} [node distance=1.5cm]
    \node(Start) [BegEnd] {Start};
    \node(PS) [startstop, below of=Start] {Craft Problem Statement};
    \node(CS) [startstop, below of=PS] {Craft Solution Strategy};
    \node(GatherData) [startstop, below of=CS] {Gather Data};
    \node(ChooseData) [startstop, below of=GatherData] {Choose and Clean Data};
    \node(Feature) [startstop, below of=ChooseData]{
      \begin{tabular}{c}
      Identify Relevant Features  \\
      Feature Engineering 
      \end{tabular}
      };
    \node(ChooseModel) [startstop, below of=Feature] {Choose Model};
    \node(SplitData) [startstop, below of=ChooseModel] {Split Data Train/Test};
    \node(TrainModel) [startstop, below of=SplitData] {Train Model};
    \node(TestModel) [focus, below of=TrainModel] {Test Model};
    \node(EP) [startstop, below of=TestModel] {Evaluate Model Performance};
    \node(DMG) [draw, below of=EP]{
      \begin{tabular}{c}
      Model approach  \\
      still believed \\
      adequate
      \end{tabular}
      };
    \node(DMF) [draw, right of=TrainModel, xshift=2cm]{
      \begin{tabular}{c}
      Model \\
      performance \\
      acceptable
      \end{tabular}
      };
    \node(End) [BegEnd, right of=DMF,xshift=1cm] {End};  
    \node(blank) [BegEnd, left of=DMG, xshift=-2cm] { };
    \draw [arrow] (Start) -- (PS);
    \draw [arrow] (PS) -- (CS);
    \draw [arrow] (CS) -- (GatherData);
    \draw [arrow] (GatherData) -- (ChooseData);
    \draw [arrow] (ChooseData) -- (Feature);
    \draw [arrow] (Feature) -- (ChooseModel);
    \draw [arrow] (ChooseModel) -- (SplitData);
    \draw [arrow] (SplitData) -- (TrainModel);
    \draw [arrow] (TrainModel) -- (TestModel);
    \draw [arrow] (TestModel) -- (EP);
    \draw [arrow] (EP) -- (DMG);
    \draw [-] (DMG) to node[anchor=south] {no} (blank.center);
    \draw [arrow] (blank.center) |- (PS);
    \draw [arrow] (DMF) |- node[anchor=west] {no} (Feature);
    \draw [arrow] (DMG) -| node[anchor=west] {yes} (DMF);
    \draw [arrow] (DMF) -- node[anchor=south] {yes} (End);
  \end{tikzpicture}
  \caption{Test Model}
  \label{fig:workflow7}
\end{figure}
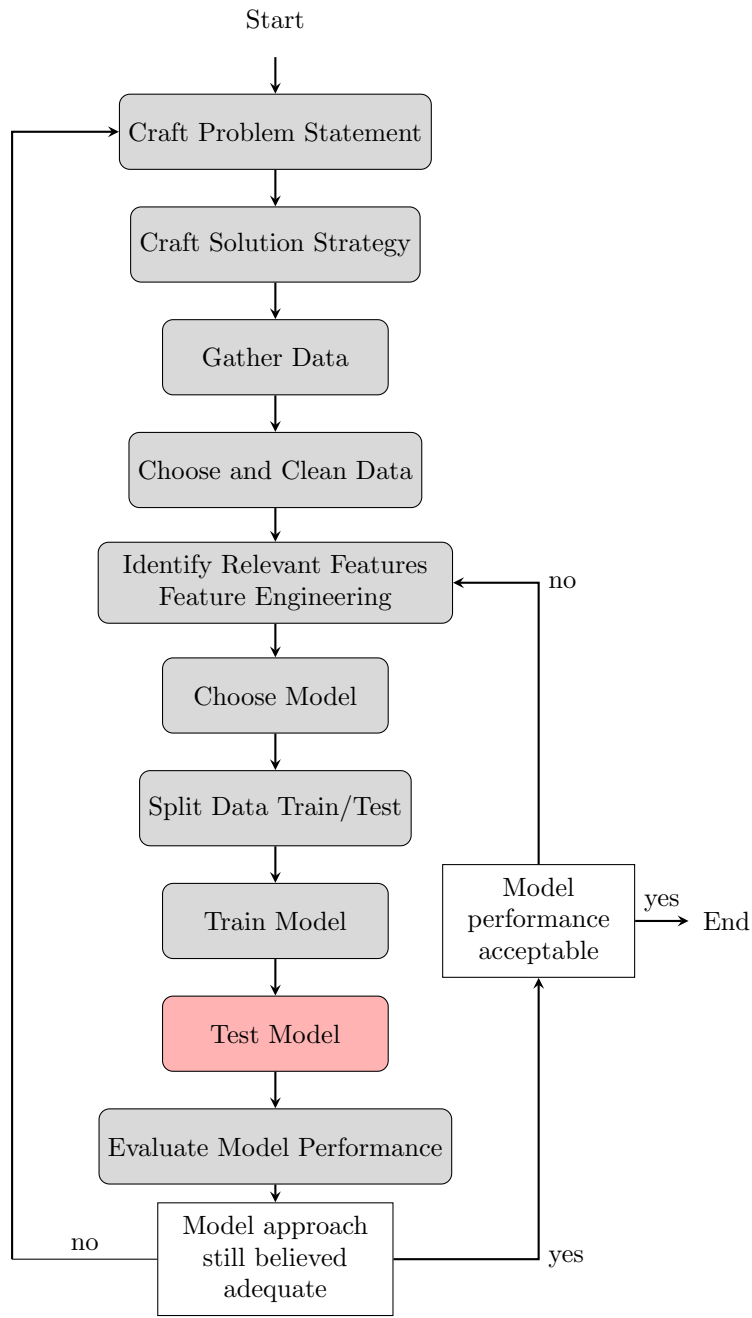

With a trained model in place, we advance to the "Test Model" stage in the ML Process flow (see figure \ref{fig:workflow7}). 

After using our training data to build a model, our next step was to supply our model with testing data to see how it performed. We had already partitioned the test data from our original dataset using the sklearn.model\textunderscore selection.train\textunderscore test\textunderscore split function, and we used this test data as input to our model. For each item in our testing dataset, the model generated a unique prediction as output. We compared the list of predictions (`y\textunderscore pred` in the code snippet \hyperref[code:makepredictions]{in figure \ref{code:makepredictions}}) against the known values (`y\textunderscore test` in the code snippet \hyperref[code:traintestsplit]{in figure \ref{code:traintestsplit}}) to understand the model's performance.
\begin{figure} [H]
\begin{lstlisting}[language=Python]
# Make predictions
y_pred = dt_classifier.predict(X_test)
\end{lstlisting}
\caption{Predicting using the test data.}
\label{code:makepredictions}
\end{figure}

With lists of predicted and expected values, we could start calculating different metrics of the model's performance.

\subsection{Evaluating the Model}

\begin{figure} [H]
  \begin{tikzpicture} [node distance=1.5cm]
    \node(Start) [BegEnd] {Start};
    \node(PS) [startstop, below of=Start] {Craft Problem Statement};
    \node(CS) [startstop, below of=PS] {Craft Solution Strategy};
    \node(GatherData) [startstop, below of=CS] {Gather Data};
    \node(ChooseData) [startstop, below of=GatherData] {Choose and Clean Data};
    \node(Feature) [startstop, below of=ChooseData]{
      \begin{tabular}{c}
      Identify Relevant Features  \\
      Feature Engineering 
      \end{tabular}
      };
    \node(ChooseModel) [startstop, below of=Feature] {Choose Model};
    \node(SplitData) [startstop, below of=ChooseModel] {Split Data Train/Test};
    \node(TrainModel) [startstop, below of=SplitData] {Train Model};
    \node(TestModel) [startstop, below of=TrainModel] {Test Model};
    \node(EP) [focus, below of=TestModel] {Evaluate Model Performance};
    \node(DMG) [draw, below of=EP]{
      \begin{tabular}{c}
      Model approach  \\
      still believed \\
      adequate
      \end{tabular}
      };
    \node(DMF) [draw, right of=TrainModel, xshift=2cm]{
      \begin{tabular}{c}
      Model \\
      performance \\
      acceptable
      \end{tabular}
      };
    \node(End) [BegEnd, right of=DMF,xshift=1cm] {End};  
    \node(blank) [BegEnd, left of=DMG, xshift=-2cm] { };
    \draw [arrow] (Start) -- (PS);
    \draw [arrow] (PS) -- (CS);
    \draw [arrow] (CS) -- (GatherData);
    \draw [arrow] (GatherData) -- (ChooseData);
    \draw [arrow] (ChooseData) -- (Feature);
    \draw [arrow] (Feature) -- (ChooseModel);
    \draw [arrow] (ChooseModel) -- (SplitData);
    \draw [arrow] (SplitData) -- (TrainModel);
    \draw [arrow] (TrainModel) -- (TestModel);
    \draw [arrow] (TestModel) -- (EP);
    \draw [arrow] (EP) -- (DMG);
    \draw [-] (DMG) to node[anchor=south] {no} (blank.center);
    \draw [arrow] (blank.center) |- (PS);
    \draw [arrow] (DMF) |- node[anchor=west] {no} (Feature);
    \draw [arrow] (DMG) -| node[anchor=west] {yes} (DMF);
    \draw [arrow] (DMF) -- node[anchor=south] {yes} (End);
  \end{tikzpicture}
  \caption{Evaluate Model}
  \label{fig:workflow8}
\end{figure}
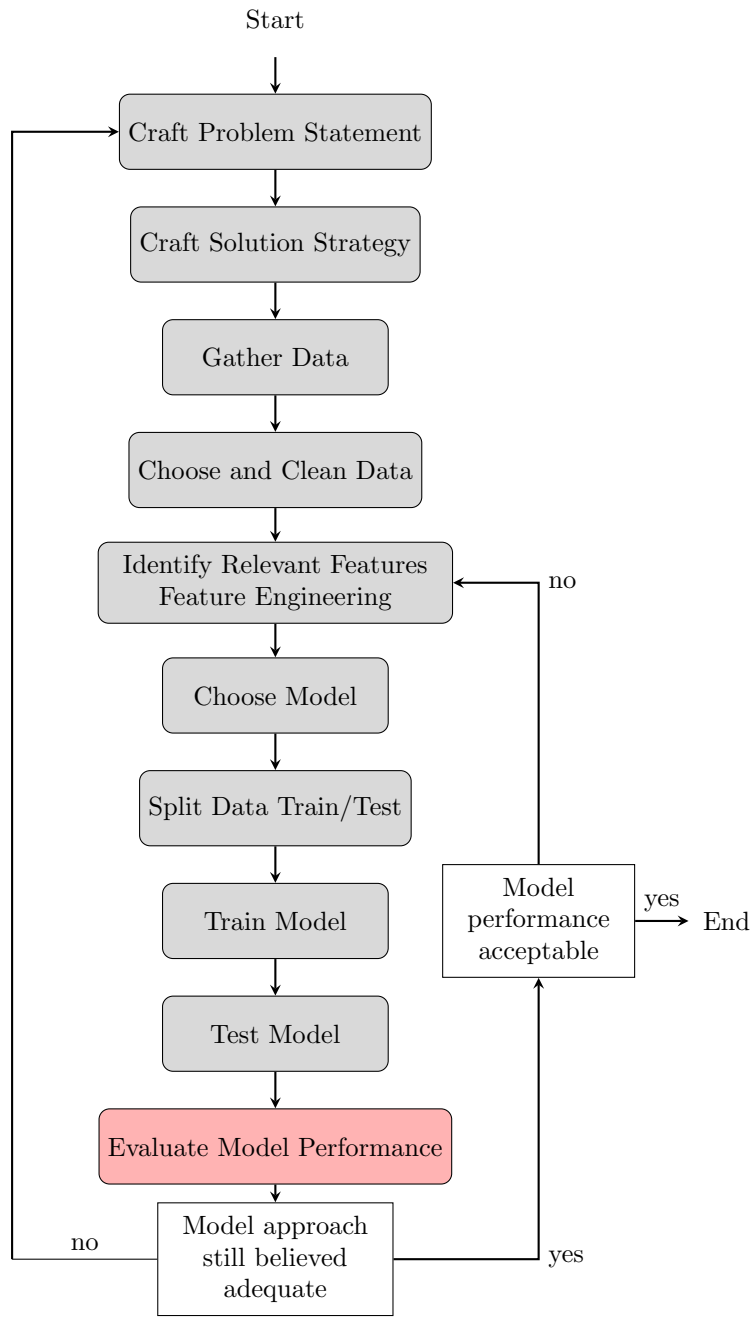

After testing the model, we can move on to the "Evaluate Model Performance" stage in the ML Process flow (see figure \ref{fig:workflow8}). 

There are multiple ways to judge a model's performance. Since we were focusing on classification problems, we looked at accuracy, precision, recall, and F\textunderscore {1}-score in the sections below. However there may be useful evaluation metrics for other types of problems (such \href{https://en.wikipedia.org/wiki/Mean_squared_error}{Mean Squared Error} for Regression or \href{https://en.wikipedia.org/wiki/Silhouette_(clustering)}{Silhouette Scoring} for Clustering).

When considering a model's results, it is important to keep the business objectives in mind. As we will cover below, there are numerous methods and metrics to evaluate a model. No model will be perfect, and sometimes you may need to make trade-offs that best align with the business's goals.

\subsubsection{Confusion Matrices and Classification Reports}

There our four possible outcomes for when a model is making a binary classification, like in our use cases.
\begin{itemize}
    \item True Positive (TP): When the model correctly predicts the positive case. In our examples, this may be correctly identifying a bad fix.
    \item True Negative (TN): When the model correctly predicts the negative case. In our examples, this may be correctly identifying that there is not a bad fix.
    \item False Positive (FP): When the model incorrectly predicts the positive case. In our examples, this may be incorrectly predicting a bad fix for code that in actuality has no issues.
    \item False Negative (FN): When the model incorrectly predicts the negative case. In our examples, this may be incorrectly predicting a that a bad fix does not contain any issue.
\end{itemize}

With these four outcomes in mind, we used Confusion Matrices and Classification reports to view a model's performance.

\label{subsec:confusionmatrices}
A Confusion Matrix compares the actual values with the predicted values, giving us the counts of True Positives, True Negative, False Positive, and False Negative results. This is visualized in figure \ref{code:ConfusionMatrix}:
\begin{figure} [H]
\begin{lstlisting}
[[ True_Negatives  False_Positives]
 [ False_Negatives True_Positives ]]
\end{lstlisting}
\caption{Confusion matrix.}
\label{code:ConfusionMatrix}
\end{figure}

Leveraging the scikit-learn packages, we created a Confusion Matrix displaying our data with the following Python code snippet in figure \ref{code:DisplayConfusionMatrix}:
\begin{figure} [H]
\begin{lstlisting}[language=Python]
from sklearn.metrics import confusion_matrix

# Display the confusion matrix for the test data predictions
print(confusion_matrix(y_test, y_pred))
\end{lstlisting}
\caption{Displaying the confusion matrix}
\label{code:DisplayConfusionMatrix}
\end{figure}

In the output below, we see that displayed Confusion Matrix provides a lot of immediate value. In this example, there were 2842 correct predictions (1307 True Positives plus 723 True Negatives) with 812 incorrect predictions (595 False Positives plus 217 False Negatives).

Running the above code will result in the following output (in figure \ref{exp:ConfusionMatrix}):
\begin{figure} [H]
\begin{lstlisting}
[[ 723  595]
 [ 217 1307]]
\end{lstlisting}
\caption{Example confusion matrix.}
\label{exp:ConfusionMatrix}
\end{figure}

A Classification Report shows the overall \hyperref[subsec:accuracy]{accuracy}, as well as the \hyperref[subsec:precisionrecall]{precision, recall}, and \hyperref[subsec:f1score]{f-scores} for each class, metrics that we will discuss in more detail below. We found classification reports can be useful when trying to determine if there were certain areas in which the model was excelling or failing in.  Figure \ref{code:DisplayClassReport} is an example of using python classification$\_$report function.
\begin{figure} [H]
\begin{lstlisting}[language=Python]
from sklearn.metrics import classification_report

# Display the confusion matrix for the test data predictions
print(classification_report(y_test, y_pred))
\end{lstlisting}
\caption{Displaying the classification report}
\label{code:DisplayClassReport}
\end{figure}

Figure \ref{code:example_classification_report} shows the classification report created by the above code. In our use case, we had two classes (true and false), and the metrics for each class can be use to evaluate performance.

\begin{figure} [H]
\begin{lstlisting}
              precision    recall  f1-score   support

        true       0.77      0.55      0.64      1318
       false       0.69      0.86      0.76      1524

    accuracy                           0.71      2842
   macro avg       0.73      0.70      0.70      2842
weighted avg       0.73      0.71      0.71      2842
\end{lstlisting}
\caption{Example classification report.}
\label{code:example_classification_report}
\end{figure}

\subsubsection{Accuracy}
\label{subsec:accuracy}

After creating an initial model and supplying it with test data, the first thing we wanted to see was the model's accuracy. In this context, accuracy means the percentage of correct predictions out of all predictions.

\[ Accuracy = \frac{True Positives + True Negative}{Total Predictions} \]

In this example (figure \ref{code:DisplayModelAccuracy}) we will use scikit-learn's accuracy\textunderscore score function passing in y\textunderscore test defined in \hyperref[code:traintestsplit]{figure \ref{code:traintestsplit}} as the list of predictions and y\textunderscore pred defined in \hyperref[code:makepredictions]{figure \ref{code:makepredictions}} as the known values.
\begin{figure} [H]
\begin{lstlisting}[language=Python]
# Display the accuracy
acc = metrics.accuracy_score(y_test, y_pred)
print(f"Accuracy: {acc}")
\end{lstlisting}
\caption{Displaying model accuracy.}
\label{code:DisplayModelAccuracy}
\end{figure}

Running the above code will result in the following output (in figure \ref{code:AccuracyOutput}):
\begin{figure} [H]
\begin{lstlisting}
Accuracy: 0.7142857142857143
\end{lstlisting}
\caption{Example accuracy output.}
\label{code:AccuracyOutput}
\end{figure}

The first time that we created a model and displayed its accuracy on the testing data was for our project trying to predict if a new code fix was bad. When our original model had a 95\%+ accuracy, we were ecstatic, believing our model was working very well right away. It was a bit deflating to learn that accuracy is not always the best indicator of model performance. We came to realize that we had significant fewer samples of "bad fixes" than "good fixes" (something that is good for our development and test teams but is bad for the context of this machine learning project), and therefore, if the model only answered "good fix" regardless of the data supplied to it, its accuracy was going to be very high. In order to understand how the model was actually performing, we needed to analyze other metrics.

\subsubsection{Precision and Recall}
\label{subsec:precisionrecall}

Our problem was that our model was always predicting "good fix," and though the overall accuracy was high, we noticed that the precision and recall metrics were poor. These two metrics lead us to the realization that our initial model needed adjustments. It is often a balancing act when trying to have good values for precision and recall, and trade-offs often need to be made. It is necessary to remember the objective and business value for the machine learning project when considering whether precision or recall is more important.

Precision answers the question: "When the model predicts 'good fix,' how often is it correct?"

\[ Precision = \frac{True Positives}{True Positives + \textbf{\textit{False Positives}}} \]

Recall, on the other hand, answers the question: "How often does the model correctly identify 'good fixes'?"

\[ Recall = \frac{True Positives}{True Positives + \textbf{\textit{False Negatives}}} \]

These two questions sound similar, but there is a distinct difference. In fact, the numerators in each formula are the same, but each has a different denominator.

\begin{example} \label{examp:precision_recall}
    Let's assume that we build a model to predict True or False, and that we wanted this model to predict against a dataset that had 10 Trues and 90 Falses. If our model always guesses True, it will have an accuracy of 10\%. Additionally, it will correctly identify all True items, and therefore have a recall of 100\%. However, the precision of the model for True values will only be 10\%.
\end{example}

In our use case, we decided that it was better for more "bad fixes" to slip through than to apply unnecessary overhead on the development team when reviewing code for a problem that never existed, so we opted for better recall.

\subsubsection{F\textunderscore {1} - score}
\label{subsec:f1score}

The F\textunderscore {1}-score is a metric that represents both the \hyperref[subsec:precisionrecall]{precision and recall}. F\textunderscore {1}-score is calculated taking the \href{https://en.wikipedia.org/wiki/Harmonic_mean}{harmonic mean} of the precision and recall for a class, with results ranging from zero to one. To measure central tendency, harmonic mean is one measurement used, representing the middle of a set of numbers by calculating the arithmetic mean of the numbers' reciprocals (1/xi) and then reciprocates the result (for more info, see \href{https://machinelearningmastery.com/arithmetic-geometric-and-harmonic-means-for-machine-learning/}{"Arithmetic, Geometric, and Harmonic Means for machine learning"}). While both precision and recall should be analyzed individually, the F\textunderscore {1}-score gives a single, combined measurement of both metrics.

An easy way to see the F\textunderscore {1}-score is by displaying the Classification Report (covered in section \ref{subsec:confusionmatrices}.

One important lesson we learned when discussing model results was that there were different ways of viewing the same data. When we examined our models' results for predicting when a field defect's fix would have an error in it, one member was happy with the same results another member was disappointed by. The disappointed member saw that of 30 "bad fixes," the model only identified 7 of them. In this person's eyes, the majority of these cases slipped by undetected. The happy member saw that when the model predicted "bad fix," it was correct 7 out of 8 times. In this person's eyes, the model was able to detect problems while introducing very little overhead of wasted work. This was an example of one person examining the precision while the other was examining the recall. It is important to analyze all the performance metrics of a model and compare its results to the business objective. Depending on that business objective, it may be better to improve the recall at the cost of precision or vice versa.

\newpage
\subsection{Testing Other Models}

After seeing the results of the Decision Tree Classifier, we wanted to know how other algorithms and types of models would perform with our data. We knew that we wanted to try using a Random Forest Classifier, since this algorithm uses sets of Decision Trees and this seemed like a logical extension. Other than that, we did not have any particular algorithms in mind, so chose a handful of some common ones. In addition to the Decision Tree Classifier, we built four new models: a Random Forest Classifier, a Naive Bayes Classifier, and two Support Vector Machines (one with an Radial Basis Function kernel and one with a polynomial kernel).

Since all the work preparing the data was already completed, it was straightforward training these new types of models. The scikit-learn packages provided Python libraries for each type of model which were all basically the same. This let us copy and paste our Decision Tree Classifier code, making minor changes to swap between types of models. The below example (in figure \ref{code:FittingModel}) shows how we built these different models using the same training data we used with the Decision Tree.
\begin{figure} [H]
\begin{lstlisting}[language=Python]
from sklearn.ensemble import RandomForestClassifier
from sklearn.naive_bayes import GaussianNB
from sklearn.tree import DecisionTreeClassifier
from sklearn.svm import SVC

# Create a train a Decision Tree Classifier
dt_classifier = DecisionTreeClassifier()
dt_classifier.fit(X_train, y_train)

# Create a train a Random Forest Classifier
rfc = RandomForestClassifier(warm_start=True, n_estimators=10)
rfc.fit(X_train, y_train)

# Create a train an SVC with a Radial Basis Function kernel
rbf_classifier = SVC(kernel='rbf', C = 1).fit(X_train, y_train)

# Create a train an SVC with a polynomial kernel
poly_classifier = SVC(kernel='poly', C = 1).fit(X_train, y_train)

# Create a train a Gaussian Naive Bayes Classifier
gnb = GaussianNB()
gnb.fit(X_train, y_train)
\end{lstlisting}
\caption{Fitting the models}
\label{code:FittingModel}
\end{figure}

While the inputs for each type of model were the same, some models had different outputs from one another. It was easy to see the feature importance metrics for Decision Tree Classifier and Random Forest Classifier, but it is actually impossible to get this information for our Support Vector Machines (SVCs). However, after training and testing, we always had a Confusion Matrix and a Classification Report, both of which helped us compare the performance of our models.

\subsection{Ensembles}

As we reviewed the prediction results for each model, we realized that there were a lot of variance, as some models were better at predicting against certain types of data than others. We also found that while certain models were better at identifying all the samples we cared about (such as the bad code fixes or invalid defects), other models were better at reducing the overhead of False Positives (again, \hyperref[subsec:precisionrecall]{Precision vs. Recall}). In order to leverage the individual strengths of each model, we decided to use an \href{https://en.wikipedia.org/wiki/Ensemble_learning}{ensemble}.

Using an ensemble is a common way to combine multiple algorithms together to produce a single prediction or result. We decided to use a Voting Classifier, which was easily implemented in Python by using scikit-learn's \href{https://scikit-learn.org/stable/modules/generated/sklearn.ensemble.VotingClassifier.html}{VotingClassifier} class from the \href{https://scikit-learn.org/stable/modules/classes.html#module-sklearn.ensemble}{ensemble} module. With a Voting Classifier, our individual models would still each make predictions. However, each models prediction of the class would be counted as a vote. In our example below (in figure \ref{code:CreateEnsemble}), we specified a voting rule of "hard," meaning that the majority vote wins.
\begin{figure} [H]
\begin{lstlisting}[language=Python]
from sklearn.ensemble import VotingClassifier

# Create a train a VotingClassifier using all the previous models
ensemble_classifier = VotingClassifier(estimators=[('DT', dt_classifier), ('RFC', rfc), ('gnb', gnb)], voting='hard')
ensemble_classifier.fit(X_train, y_train)
\end{lstlisting}
\caption{Created an ensemble}
\label{code:CreateEnsemble}
\end{figure}

By implementing an ensemble, we found that our overall performance had a humble increase against our data. Our Voting Classifier did not magically fix all problems but it did perform better and more consistently. Previously, we had found that our models' prediction accuracies would vary greatly depending on how the random sampling of training and testing data was done (we had not yet discovered \hyperref[subsec:crossvalidation]{Cross Validation}, a way of training against subsets of data). By leveraging the strengths of the individual models, the ensemble helped us improve our overall performance and consistency.

\section{Next Steps}

While our initial journey gave us great experience into the world of machine learning, there were a few topics and strategies that we could have tried to improve results. Many of these are areas we are planning to learn about, experiment with, and hopefully implement.

\subsection{Hyperparameters}

When building our models, we focused on adjusting the data. This is a valid and valuable practice, but it is not the only way to improve a model's performance. In addition to working with the data, the model itself could also be tuned through the altering of \href{https://en.wikipedia.org/wiki/Hyperparameter_(machine_learning)}{hyperparameters}. An example of a hyperparameter for a Decision Tree Classifier is the maximum tree depth. One reason to reduce the maximum tree depth could be to reduce the model's overall complexity and potentially the overall number of splits.

\subsection{Cross Validation}
\label{subsec:crossvalidation}

When building a model, noise and variance in the training data could lead to poor results. In order to assure our model was properly looking at the right patterns in the data, \href{https://scikit-learn.org/stable/modules/generated/sklearn.model_selection.cross_validate.html}{Cross Validation} could be used. As we were training our models, we found that the model's performance greatly varied depending on how the training and testing data was split. This is a big problem and it is important to build a model that can handle all samples well. K-Fold Cross Validation is one technique to combat this problem. The original dataset is split into k groups (called folds), where one fold is the testing dataset while the remaining folds make up the training dataset. The process is repeated until each fold is used for both training and testing.

\subsection{Model Deployment}

While building and testing our model was done locally, we would need to deploy our model and expose interfaces in order for a larger team to get value out of it. There are various products and tools that build an \href{https://en.wikipedia.org/wiki/MLOps}{MLOps} pipeline.

For our examples, we may want to create a pipeline that runs when a new software defect record is created. The data provided in the defect could be supplied to a machine learning model so a prediction could be made. If this information could be presented back to the defect creator, the creator could use this information in some manner (such a reviewing for errors, re-prioritizing, or whatever is appropriate).

\section{Appendix}

\subsection{Acknowledgements}

Finally, we would like to extend a heartfelt "thank you" to...
\begin{itemize}
    \item Eitan Farchi: For the expertise, education, and answers you so frequently shared with us.
    \item James O'Connor: For the guidance and encouragement to try new things and create something valuable.
    \item Michael Gildein: For the ideas, examples, and soundboarding that led us to new topics and techniques.
\end{itemize}



\end{document}